\documentclass[aps,prd,twocolumn,a4paper,superscriptaddress,10pt,floatfix]{revtex4-1}
\usepackage{graphicx}
\usepackage{multirow}
\usepackage{latexsym}
\usepackage{hyperref}
\usepackage[british]{babel}
\usepackage{amsmath,amsmath}
\usepackage{color}
\usepackage[dvipsnames]{xcolor}
\usepackage{amsmath}
\begin{document}
\def\cjm#1{{\color{ForestGreen} \sf #1}}
\def\kng#1{{\color{RawSienna} \sf #1}}
\def\elk#1{{\color{Aquamarine} \sf #1}}
\newcommand{\be}{\begin{equation}}
\newcommand{\ee}{\end{equation}}
\newcommand{\bq}{\begin{eqnarray}}
\newcommand{\eq}{\end{eqnarray}}

\title{Probing fundamental physics using compact astrophysical objects}
\author{E.-A. Kolonia}
 \email{ekolonia@perimeterinstitute.ca}
 \affiliation{Department of Physics, University of Patras, Patras, Rio, 26504, Greece}
 \affiliation{Perimeter Institute for Theoretical Physics, 31 Caroline Street North, Waterloo, ON, N2L 6B9, Canada}
\author{C. J. A. P. Martins}
\email{Carlos.Martins@astro.up.pt}
\affiliation{Centro de Astrof\'{\i}sica da Universidade do Porto, Rua das Estrelas, 4150-762 Porto, Portugal}
\affiliation{Instituto de Astrof\'{\i}sica e Ci\^encias do Espa\c co, Universidade do Porto, Rua das Estrelas, 4150-762 Porto, Portugal}
\author{Konstantinos N. Gourgouliatos}
\email{kngourg@upatras.gr}
 \affiliation{Department of Physics, University of Patras, Patras, Rio, 26504, Greece}
\date{\today}

\begin{abstract}
It is well known that alternative theories to the Standard Model allow---and sometimes require---fundamental constants, such as the fine-structure constant, $\alpha$, to vary in spacetime. We demonstrate that one way to investigate these variations is through the Mass-Radius relation of compact astrophysical objects, which is inherently affected by $\alpha$ variations. We start by considering the model of a polytropic white dwarf, which we perturb by adding the $\alpha$ variations for a generic class of Grand Unified Theories. We then extend our analysis to neutron stars, building upon the polytropic approach to consider more realistic equations of state, discussing the impact of such variations on mass-radius measurements in neutron stars. We present some constraints on these models based on current data and also outline how future observations might distinguish between extensions of the Standard Model.
\end{abstract}
\maketitle

\section{\label{sec:intro}Introduction}

The quest to understand the fundamental laws of physics has driven scientific inquiry for centuries. From the classical mechanics of Newton to the quantum mechanics and relativity of the XX century, physicists have continually sought to unify and explain the physical mechanisms and forces governing the universe. A wide and fast-growing range of observational data has so far validated the Standard Model of particle physics and that of cosmology ($\Lambda$CDM), but in parallel there is growing evidence that this standard model is incomplete---or, pragmatically, a simple convenient approximation to a more fundamental one. One of the most ambitious goals in this attempt is the formulation of Grand Unified Theories (GUTs), which aim to merge the electromagnetic, weak, and strong nuclear forces into a single framework \cite{Langacker}. Despite advances in theoretical physics, experimental verification of GUT predictions is still limited \cite{PDG}. 

Among the testable consequences of the GUT class of models are spacetime variations of nature's fundamental constants (such as the fine-structure constant, $\alpha$). Recent observational developments---see \cite{ROPP,Uzan} for detailed reviews---have significantly improved the precision and accuracy of these searches, making them a competitive probe of fundamental cosmology. Compact objects such as white dwarfs and neutron stars serve as natural laboratories for probing these deep questions \cite{Shapiro,Ozel}, and in particular they are sensitive to such variations. For example, constraints on both $\alpha$ and the proton-to-electron mass ratio $\mu$ have been obtained on the surface of white dwarfs \cite{Berengut,Bagdonaite}, which enables constraints on environmental dependencies in these models \cite{Olive}.

Another example, which is the focus of the present work, is the mass-radius relation of white dwarfs and neutron stars. This encodes much of the underlying physics of these objects, and is therefore expected to be affected in extensions of the standard models of cosmology and/or particle physics. For the simpler case of white dwarfs, this relation has already been considered as a test of modified gravity scenarios \cite{Jain} and also, under a polytropic approximation, as a test of unification scenarios \cite{Magano}. In what follows we update and extend the earlier analyses, both by including neutron stars and by going beyond the polytropic approximation.

The structure of the rest of this work is as follows. We start in Sect.~\ref{sec:theory} by presenting a brief overview of the relevant theoretical background behind varying couplings and unification (for which we consider three representative GUT class models), as well as the properties of compact objects. In Sect.~\ref{sec:wd} we investigate a polytropic white dwarf, both to provide a simple illustration of our formalism and to gain some intuition on the impact of these models on the mass-radius relation. We then move to the case of neutron stars in Sect.~\ref{sec:ns}, for which we build upon the polytropic approximation to explore more realistic choices of the equation of state (EoS), specifically considering three representative examples. Finally, we present our conclusions in Sect.~\ref{sec:concl}.

\section{\label{sec:theory}Theoretical background}

There is a plethora of astrophysical evidence---including indications for dark matter, neutrino masses, inflation, and the size of the baryon asymmetry in the universe---supporting the view that some still undiscovered physics beyond the current model is relevant for the evolution of the universe. On the theory side, one can consider extensions or even alternative models to the standard one, at least for the phenomenological purpose of ascertaining the degree to which fundamental laws and symmetries, such as the Einstein Equivalence Principle or Lorentz invariance, are allowed to be violated by terrestrial experiments and astrophysical observations.

Scalar fields are arguably the best-motivated conceptual extension to the standard cosmological paradigm, and they are essential components of many classes of theories, due to their simplicity combined with their effectiveness. They can preserve Lorentz invariance while simultaneously taking a vacuum expectation value, which is not possible for other types of fields. Importantly, they have been experimentally proven to exist: the discovery of the Higgs particle \cite{ATLAS,CMS} contributed to their popularity among theorists.

Among the generic measurable effects of dynamical scalar fields are spacetime variations of nature's fundamental constants. It is commonly known that dimensionless couplings \textit{run} with energy \cite{PDG}, but in many extensions of the Standard Model, including those where scalar fields couple to the rest of the Lagrangian (specifically, to its electromagnetic sector), such couplings also \textit{roll} in time \cite{1998PhRvL..81.3067C} and \textit{ramble} in space \cite{OlivePospelov}. Constraining these variations is a powerful probe of fundamental cosmology \cite{ROPP,Uzan}. The present work considers possible variations of the fine-structure constant, defined as
\begin{equation}
    \frac{\Delta\alpha}{\alpha}(X)=\frac{\alpha(X)-\alpha_0}{\alpha_0} 
\end{equation}
with $\alpha_0$ being the local present-day value of $\alpha$ and $X$ generically denoting another spacetime location. If one is considering possible time dependencies this would be redshift, but in the present work this will denote the value in the compact objects under study.

The details of the expected $\alpha$ variations depend on the models being considered. We consider a broad class of GUTs \cite{Coc} where the weak scale is determined by dimensional transmutation, relative variations of all the Yukawa couplings are the same, and the variation of the couplings is driven by a dilaton-type scalar field \cite{Campbell}. In addition to the previously mentioned related work on white dwarfs \cite{Magano}, this class of models has also been assumed in other theoretical studies of solar-type stars \cite{Vieira}, Population III stars \cite{Ekstrom} and (with some simplifications) neutron stars \cite{Perez}.

These models are phenomenologically useful because they enable a self-consistent description of variations in all relevant model parameters which can be described in terms of only three of them: $\alpha$ itself, and two other dimensionless parameters, denoted $R$ and $S$ and defined as
\bq
\frac{\Delta\Lambda}{\Lambda}&=&R\frac{\Delta\alpha}{\alpha}+(Electroweak\; terms)\\
\frac{\Delta v}{v}&=&S\frac{\Delta h_i}{h_i}\,,
\eq
where $\Lambda,\;v,\;h_i$ are the QCD scale, Higgs vacuum expectation value, and Yukawa couplings respectively. The parameters $R$ and $S$ are related to the energy scale at which unification is assumed to occur, with the first being related to QCD sector and the second to electroweak physics. Of particular relevance for what follows is the fact that in this model class a change in $\alpha$ will impact particle masses. For the electron mass this immediately follows from the change in the Higgs vacuum expectation value and the Yukawa couplings (bearing in mind that $m_e=h_ev/\sqrt{2}$). For composite particles such as nucleons an analogous but lengthier argument follows or, more phenomenologically, one may recall that that the masses of the proton and neutron include electromagnetic correction terms (which depend on the value of $\alpha$) \cite{Gasser}.

Phenomenologically, each choice of a pair of values for $(R,S)$ corresponds to one such model in this class. In our analysis we use three specific GUTs models, intended as representatives of the wider $(R,S)$ parameter space. These are
\begin{itemize}
    \item The \emph{Unification} model, with $(R=36, S=160)$, has arguably the most typical parameter values in this model class \cite{Coc};
    \item The \emph{Dilaton} model, with $(R=109.4, S=0)$, which draws inspiration from string-theory type scalar fields \cite{Nakashima};
    \item The \emph{UV-Cutoff} model, with $(R=-183, S=22.5)$, in which one assumes that this cutoff is cosmologically varying \cite{Lee}.
\end{itemize}

Compact objects are formed through the gravitational collapse of progenitor stars during their final evolutionary stages. Stars of up to about 9 $M_{\odot}$ become Carbon-, Oxygen- and sometimes Helium-rich white dwarfs, while more massive ones collapse into neutron stars \cite{2003ApJ...591..288H}. Being former cores of stars, white dwarfs and especially neutron stars are extremely dense. They are small in size (about the size of the Earth and an asteroid respectively), but have masses comparable to that of (or bigger than) the Sun. Since thermonuclear reactions are absent from their interiors, they rely on degeneracy pressure to support themselves against gravity. Degeneracy pressure for a non-relativistic electron gas is defined as $P_e=(3\pi^2)^{2/3}\hbar^2n_e^{5/3}/(5m_e)$ whereas for the relativistic gas, we alter the coefficients so that the dependence on the number density is $P_e\propto n_e^{4/3}$. We can model their dynamics and derive the Mass-Radius relation from the Tolman-Oppenheimer-Volkoff (TOV) equations \cite{Shapiro,Silbar,Sagert}
\begin{align}
     \frac{dp}{dr}&=-\frac{G\rho(r)M(r)}{r^2}\nonumber
     {}\\
     &\times \left[1+\frac{p(r)}{\epsilon(r)}\right]\left[1+\frac{4\pi r^3p(r)}{M(r)c^2}\right]\left[1-\frac{2GM(r)}{c^2r}\right]^{-1}
     \label{TOV1}\\
    \frac{dM}{dr}&=4\pi r^2\rho(r)\,,
    \label{TOV2}
\end{align}
together with an EoS. The only analytic mass-radius relation is that of a polytrope $p=K\epsilon^\frac{n+1}{n}$ where $K$ is the polytropic coefficient and $n$ the polytropic index:
\begin{align}
    M&=4\pi c^{(2n+2)/(n-1)}\left(\frac{(n+1)K}{4\pi G}\right)^{n/(n-1)} \nonumber\\ 
    &\times \xi_1^{(n-3)/(1-n)}\xi_1^2|\theta'(\xi_1)|R^{(3-n)/(1-n)}
    \label{MR polytrope}
\end{align}
where $\xi_1$ corresponds to the first zero of the dimensionless density $\theta$ \cite{Chand}.

Our goal is to study the impact of non-standard values of $\alpha$ on the mass-radius relation for compact objects (and how sensitive this relation is to these values), using as examples the three phenomenological models introduced above. For simplicity we will assume that the value of $\alpha$, although different from the local one, is constant in each compact object---in other words, it does not vary as a function of the object's radius, which would be a further environmental dependence. A brief discussion of that scenario can be found in the appendix of \cite{Magano}.

\section{\label{sec:wd}Polytropic White Dwarfs}

Our study of the impact of varying fundamental constants on the mass-radius relation follows a perturbative approach,  since any such variations are anticipated to be small. The first step in our analysis is to substitute the particle masses with their respective dimensionless couplings
\begin{equation}
    \alpha_i=\frac{Gm_i^2}{\hbar c}\,;
    \label{ai=f(mi)}
\end{equation}
we make the metrological choice that particle masses vary, while Newton's gravitational constant does not. (The alternative would be to allow the Planck mass---and therefore G---to vary, with particle masses kept constant, which would be physically equivalent to our choice.) As a result, mass variations for electrons and nucleons can be written \cite{Coc,Magano}
\bq
    \frac{\Delta\alpha_e}{\alpha_e}&=&2\frac{\Delta m_e}{m_e}=(1+S)\frac{\Delta\alpha}{\alpha}\\
    \frac{\Delta\alpha_N}{\alpha_N}&=&2\frac{\Delta m_N}{m_N}=2[0.8R+0.2(1+S)]\frac{\Delta\alpha}{\alpha}\,.
\eq
These perturbations need to be included in the TOV Eqs. (\ref{TOV1}, \ref{TOV2}), as well as the Mass-Radius relation (Eq.~\ref{MR polytrope}). To accomplish this, we shift to Planck units (which we will denote with $*$ indices), since they remain unchanged, and substitute masses with couplings using Eq.~\ref{ai=f(mi)}.

Starting with the mass-radius relation, we note that only $K$ has a particle mass dependency. In the non-relativistic case, this becomes
\begin{equation}
    M_*^{1/3}R_*=Q_{NR}\alpha_e^{-1/2}\alpha_N^{-5/6},
\end{equation}
where $Q_{NR}$ is a dimensionless constant which we do not explicitly need. Then, we Taylor expand for both couplings to add perturbations
\begin{equation}
    \left(\frac{M_*(\alpha)}{M_{*,0}}\right)^{1/3}\frac{R_*(\alpha)}{R_{*,0}}=1-x,
\end{equation}
where the zero subscript indicates the standard quantity, and $x$ is the perturbation dependent on the $(R,S,\Delta\alpha/\alpha)$ model
\begin{equation}
    x=\left[\frac{4}{3}R+\frac{5}{6}(1+S)\right]\frac{\Delta\alpha}{\alpha}\,.
    \label{x}
\end{equation}
Similarly in the relativistic case, we find
\begin{equation}
    M_*=Q_R\alpha_N^{-1}
\end{equation}
and therefore
\begin{equation}
    \frac{M_*(\alpha)}{M_{*,0}}=1-y
\end{equation}
with
\begin{equation}
    y=\left[\frac{8}{5}R+\frac{2}{5}(1+S)\right]\frac{\Delta\alpha}{\alpha}\,.
    \label{y}
\end{equation}

Applying the same methodology to the non-relativistic TOV equations we find
\bq
    \frac{dp_*}{dr*}&=&-O_1\frac{m_*p_*^{3/5}}{r_*^2}\alpha_e^{3/10}\alpha_N^{1/2}\left[1+\frac{3}{5}x\right]\\
    \frac{dm_*}{dr_*}&=&O_2r_*^2p_*^{3/5}\alpha_e^{3/10}\alpha_N^{1/2}\left[1+\frac{3}{5}x\right]\,,
\eq
while for the relativistic case (without corrections) we have
\bq
\label{relativ p dwarfs}
    \frac{dp_*}{dr*}&=&-O_3\frac{m_*p_*^{3/4}}{r_*^2}\alpha_N^{1/2}\left[1+\frac{1}{2}y\right]\\
    \frac{dm_*}{dr_*}&=&O_4r_*^2p_*^{3/4}\alpha_N^{1/2}\left[1+\frac{1}{2}y\right]
    \label{relativ m dwarfs}\,,
\eq
where $x,y$ are exactly the previously defined parameters while the $O_i$ are again dimensionless numerical constants. It is also important to note that the above perturbations do not affect the zero-order terms (the general form of the equations), so we can always revert the equations back to the original units and add the corrections. It follows that there is a clear and comparatively simple procedure for including these dimensionless corrections in previously developed numerical codes.

Finally, considering the relativistic corrections in the equation of hydrostatic equilibrium (Eqs.~\ref{TOV1}, \ref{TOV2}), it becomes clear that only the first term has an $\alpha$ dependency. As a result, it is perturbed as follows for the non-relativistic case
\begin{equation}
    1+\frac{p}{\epsilon} \xrightarrow{}1+\frac{Op_*^{2/5}}{\alpha_e^{3/10}\alpha_N^{1/2}}\left[1-\frac{3}{5}x\right]\,,
\end{equation}
while for the relativistic case we get
\begin{equation}
    1+\frac{p}{\epsilon} \xrightarrow{}1+\frac{Op_*^{1/4}}{\alpha_N^{1/2}}\left[1-\frac{1}{2}y\right]
    \label{relativ correction dwarfs}\,.
\end{equation}

\begin{figure*}
  \begin{center}
    \includegraphics[width=\columnwidth]{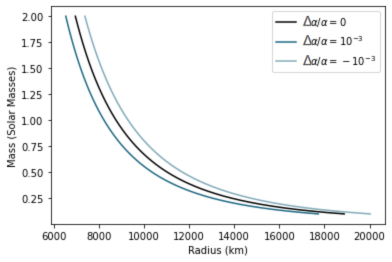}
    \includegraphics[width=\columnwidth]{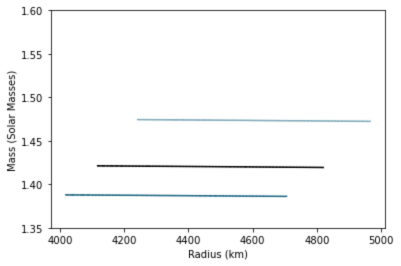}
    \includegraphics[width=\columnwidth]{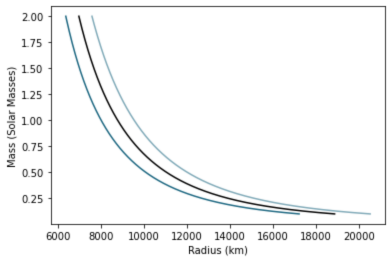}
    \includegraphics[width=\columnwidth]{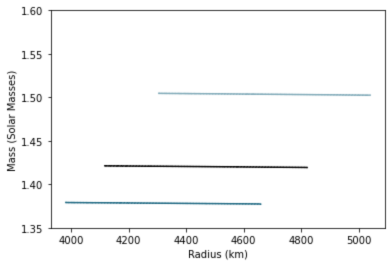}
    \includegraphics[width=\columnwidth]{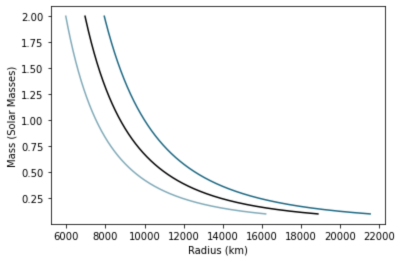}
    \includegraphics[width=\columnwidth]{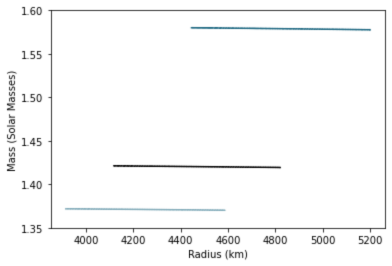}
    \caption{White dwarf Mass-Radius relation for the Unification, Dilaton and UV cutoff models (top, middle, and bottom panels respectively) in the non-relativistic (left) and relativistic (right) case for positive and negative variations of $\alpha$ as well as the standard value. Notice that the relativistic plot indicates the model's value for Chandrasekhar limit.}
    \label{fig1}
  \end{center}
\end{figure*}

We can now illustrate the effects of a variation of $\alpha$ on white dwarfs. We start by assuming a variation $\Delta\alpha/\alpha=\pm10^{-3}$ which is small enough to act as a perturbation while still having noticeable effects. We allow for both positive and negative variations, to check whether the perturbed models are symmetrical with respect to the unperturbed white dwarf. Figure \ref{fig1} compares these results for the three GUT models being considered.

In the Unification model (cf. the top panels in the figure), both plots agree on the fact that a negative $\Delta\alpha/\alpha$ increases the mass and radius of the star, while a positive variation causes the opposite effect. This can be understood since the strength of the electron degeneracy pressure, which is related to $\alpha$ as
\be
    \frac{P_e(\alpha)}{P_{e,0}}=1-\frac{1}{2}(1+S)\frac{\Delta\alpha}{\alpha}\,,
\ee
will increase for negative $\alpha$ variations as long as $S$ is non-negative (which is the case for all the models we consider, and physically expected to be the case in general). Additionally, the $(R,S)$ values affect the QCD scale (and therefore the proton and neutron masses) and the electroweak interaction respectively. In the Unification model, this effect on the electron degeneracy pressure goes in the same direction as the one from the change in the QCD scale, due to a positive $R$; a negative/positive $\Delta\alpha/\alpha$ reduces/increases the individual nucleon masses, respectively, and therefore the corresponding gravitational field. The final result is that for positive $\Delta\alpha/\alpha$ the star must therefore have a smaller mass to remain in equilibrium. For the relativistic case, it is important to note that the Chandrasekhar limit is also shifted to lower or higher masses respectively, depending on the sign of the $\alpha$ variation. It is also clear that the perturbed cases in the relativistic model are not symmetric to the unperturbed case. This occurs due to the perturbation added to one of the correction terms in the hydrostatic equilibrium equation (cf. Eq.\ref{TOV1}). 

A similar behavior can be seen in the Dilaton model case (cf. the figure's middle panels). This time the larger positive $R$ value causes larger shifts from the unperturbed case. In this model, for a negative $\alpha$ variation we observe a Chandrasekhar limit at about $1.5M_{\odot}$, significantly larger than the mass of ZTF J1901+1458, the most massive white dwarf ever detected $(1.327-1.365 M_{\odot})$ \cite{2021Natur.595...39C}. This suggests that the Dilaton model with a negative variation of $\alpha$ is more tightly constrained than the corresponding unification model, although this is compounded by the astrophysical uncertainties regarding how common the formation of white dwarfs close to the theoretical Chandrasekhar limit is, and how long these survive once formed.

The UV Cutoff models presents us with entirely different results (cf. the figure's bottom panels). The overall trends of the previous models are now reversed, with the positive $\alpha$ variation resulting in an increase in mass and radius. This is due to the negative $R$ value in this model, which reduces the QCD scale. While for a positive $\Delta\alpha/\alpha$ electron degeneracy pressure is reduced as before, this effect is counteracted and overwhelmed by the QCD-induced decrease in the nucleon masses. Therefore the comparison between the models highlights the trade-off between the impacts of a varying $\alpha$ on electrons and nucleons.

\begin{figure*}
  \begin{center}
    \includegraphics[width=\columnwidth]{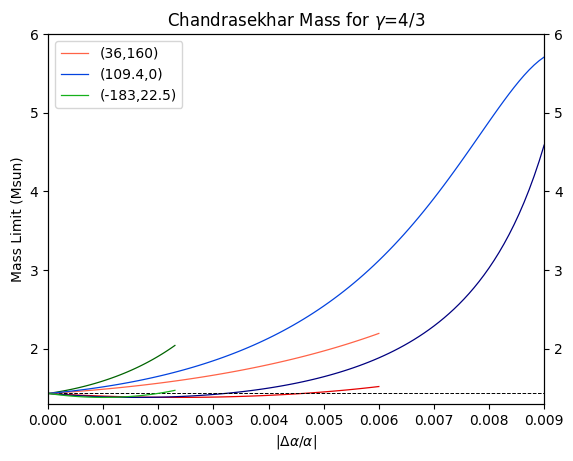}
    \includegraphics[width=\columnwidth]{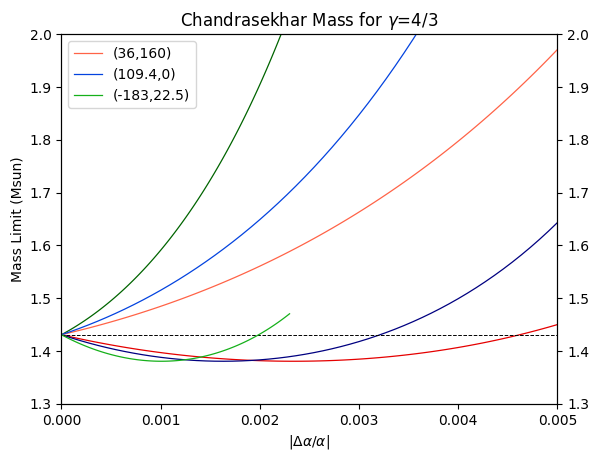}
    \caption{The Chandrasekhar mass limit as a function of the absolute value of the $\alpha$ variation for the three GUT models. Red, blue, and green colors denote the Unification, Dilaton, and UV-Cutoff models respectively, while solid and dashed lines refer to positive  and negative variations of $\alpha$ respectively. Numerical limitations, including the perturbations becoming comparable to the zeroth-order term, lead to different ranges of $\alpha$ values which can be integrated for each GUT model. The right-side panel is a zoom-in of the bottom left corner of the left-side panel, and the black dotted line shows the unperturbed Chandrasekhar limit.}
    \label{fig2}
  \end{center}
\end{figure*}

A more detailed exploration of the Chandrasekhar mass limit is shown in Figure \ref{fig2} where we now allow the value of $\Delta\alpha/\alpha$ to be a free parameter. As a consistency check, when $\Delta\alpha/\alpha$ equals $10^{-3}$ we recover the previously reported results. As we increase the absolute value of the $\alpha$ variations, the first-order term in Eqs. (\ref{relativ p dwarfs}, \ref{relativ m dwarfs}, \ref{relativ correction dwarfs}) starts becoming comparable to or even larger than the zeroth-order term. This leads to a non-symmetric behavior of the mass limit with respect to the unperturbed one, and indeed to a non-monotonic one: for a combination of $(R,S)$ values and sign of $\alpha$ variations such that the Chandrasekhar mass is larger than the standard value in the limit of small variations, this mass always increases monotonically as the magnitude of the $\alpha$ variation increases. On the other hand, for parameter combinations such that small variations make the Chandrasekhar mass smaller than the standard value, this trend is eventually reversed as the magnitude of the variation increases, and for sufficiently large variations the modified Chandrasekhar mass also increases. Specifically, the approximate ranges of $\alpha$ variations leading to values of the Chandrasekhar mass lower than the standard one in each of our three models are
\bq
0 < &\frac{\Delta\alpha}{\alpha}&\le 0.0045\quad {\rm Unification}\\
0 < &\frac{\Delta\alpha}{\alpha}&\le 0.0032\quad {\rm Dilaton}\\
-0.0020 \le &\frac{\Delta\alpha}{\alpha}&<0\quad {\rm UV Cutoff}\,.
\eq
This shows that observational limits on white dwarf masses will lead to different constraints on the various models in this class.

\section{\label{sec:ns}Neutron Stars}

We can now extend the analysis of the previous section to neutron stars. In this case, apart from the standard analytical equations that describe the interior of a compact object, there are more accurate EoS in the form of numerical tables. Given that the pressure gradient in a neutron star is significant, we should account for phase transitions as well as relativistic or hyper-relativistic phenomena near the center of the star. This can be accomplished using numerical EoS.

Here we initially work with the SLy4 model \cite{2015PhRvC..92e5803G,RADUTA2019252}, acquired from the CompOSE database \cite{2015PPN....46..633T,2017RvMP...89a5007O,2022EPJA...58..221C}. SLy4 is a stiff EoS that assumes degenerate matter comprised of nucleons and electrons. It is based on Skyrme-type interactions and predicts small neutron star radii. As a test of the sensitivity of our results to this EoS assumption, we explore two additional examples, which span most of the realistic EoS parametric space, FSU2R \cite{PhysRevC.90.045803, 2017PASA...34...65T, 10.3389/fspas.2019.00013, 10.1093/mnras/stz800} and DD2 \cite{PhysRevC.90.045803, 2010PhRvC..81a5803T, 10.1093/mnras/stz800}. FSU2R is a refined version of the FSU2 EoS, focusing on neutron-rich matter. It incorporates adjustments for nuclear symmetry energy, resulting in an intermediate stiffness, and predicts compact neutron star radii while supporting high maximum masses. DD2 provides moderate pressure at high densities, leading to larger radii than SLy4. It is also well-aligned with nuclear physics data and astrophysical constraints, supporting massive stars above $2 M_\odot$ \cite{PhysRevC.90.045803, 2017PASA...34...65T, 10.3389/fspas.2019.00013, 10.1093/mnras/stz800, 2010PhRvC..81a5803T, Carreau2019, PhysRevC.94.035804, universe7100373}. A brief comparison of the three EoS can be seen in Table \ref{tab:eos_comparison}.

\begin{table*}
\centering
\begin{tabular}{|l|c|c|c|l|}
\hline
\textbf{Model} & \textbf{Radius (1.4 $M_\odot$)} & \textbf{Max Mass ($M_\odot$)} & \textbf{L (MeV)} & \textbf{Use Case} \\ \hline
SLy4           & ~11.7 km                       & ~2.05                         & ~45              & Compact stars; soft EoS       \\ \hline
FSU2R          & ~12.4 km                       & ~2.1-2.3                      & ~47              & Balanced; neutron-rich matter \\ \hline
DD2            & ~13.2 km                       & ~2.42                         & ~55-60           & Heavy stars; stiffer EoS      \\ \hline
\end{tabular}
\caption{Comparison of the SLy4, DD2, and FSU2R equations of state for neutron stars (L refers to the symmetry energy slope parameter).\cite{Carreau2019, PhysRevC.94.035804, universe7100373} }
\label{tab:eos_comparison}
\end{table*}

As shown in the previous section, the Mass-Radius relation in the non-relativistic and relativistic limits is the following
\bq
    M_*^{1/3}R_*&\propto &\frac{1}{\alpha_e^{1/2}\alpha_N^{5/6}},\\
    M_*&\propto& \frac{1}{\alpha_N}\,.
\eq
Since we are assuming that perturbations due to varying $\alpha$ are small, we can phenomenologically interpolate between the two cases. This will enable us to introduce an effective (pressure-dependent) polytropic index. Our EoS can similarly be parameterized through an effective index (to be further discussed below), allowing us to still apply the methodology of the previous section. We must emphasize, however, that we are not assuming that the neutron star is described by a polytope. This more general relation has the form
\begin{equation}
    M_*^{n-1}R_*^{3-n}\propto \frac{1}{\alpha_e^{(3-n)/2}\alpha_N^{(1+n)/2}}\,.
\end{equation}
The dependence of the various terms on $n$ is determined by the need to have a dimensionally correct relation and recovering the two limits; the proportionality factor will depend on the equation of state, which we will discuss below. It follows that in our formalism the perturbed case should be
\begin{equation}
    \left(\frac{M_*(\alpha)}{M_{*,0}}\right)^{(n-1)}\left(\frac{R_*(\alpha)}{R_{*,0}}\right)^{(3-n)}=1-z
\end{equation}
where
\begin{equation}
    z=\left[\frac{4}{5}(n+1)R+\frac{17-3n}{10}(1-S)\right]\frac{\Delta\alpha}{\alpha}\,,
    \label{z}
\end{equation}
from which we can indeed recover the two previous limits. We emphasize that this approach is purely phenomenological, meaning that the effective polytropic index $n$ does not necessarily have a straightforward physical meaning. Still, this is not significantly different from having an EoS provided as a numerical table. Alternative (but still purely phenomenological) approaches to this issue have been considered in \cite{Read,Raithel,Suleiman,Lindblom,Reed}. In what follows we explore the implications of these choices for the Mass-Radius relation as a whole.

From the TOV equations, the EoS is encoded in the parameter $K$, whose behaviour we also need to parameterize. This can be done by demanding that it always has the correct physical dimensions, namely $m^{1/n}s^{2/n}/(kg)^{1/n}$, in addition to the two limits, leading to the unique consistent option
\begin{equation}
    K\propto\frac{\hbar^{3/n}}{m_e^{-1+3/n}m_N^{1+1/n}c^{5/n}}\,.
\end{equation}
Then the TOV equations become
\bq
    \label{dpdr perturbed ns}
    \frac{dp_*}{dr_*}&=&-O_5\frac{m_*p_*^{\frac{n}{1+n}}}{r_*^2}\alpha_e^{\frac{3-n}{2(1+n)}}\alpha_N^{1/2}(1+A)\\
    \frac{dm_*}{dr_*}&=&O_6r_*^2p_*^{\frac{n}{1+n}}\alpha_e^{\frac{3-n}{2(1+n)}}\alpha_N^{1/2}(1+A),
    \label{dmdr perturbed ns}
\eq
where again the $O_i$ are numerical constants whose values are not explicitly needed for what follows, and
\begin{equation}
    A=\left[\frac{4}{5}R+\frac{17-3n}{10(1+n)}(1+S)\right]\frac{\Delta\alpha}{\alpha}\,;
    \label{A}
\end{equation}
one immediately notices that $z=(1+n)A$. Finally, for the first of the relativistic corrections in the TOV equation, we have
\begin{equation}
    1+\frac{p}{\epsilon}\xrightarrow{}1+\frac{Op_*^{\frac{n}{1+n}}}{\alpha_e^{\frac{3-n}{2(1+n)}}\alpha_N^{1/2}}(1-A)\,.
\end{equation}
where $O$ is again a numerical constant that varies for different $n$ values.
It is important to emphasize that now the perturbation terms (Eqs. \ref{z}, \ref{A}) are not only dependent on the GUT model and the $\alpha$ variation: there is an additional dependency on the effective polytropic index.

The main issue with numerical EoS in our analysis is that, so far, we have only perturbed the model using analytical EoS expressions. More specifically, Eqs. (\ref{dpdr perturbed ns}, \ref{dmdr perturbed ns}) effectively assume the use of a polytrope, or at least some EoS parametrized as a polytrope. Consequently, in order to be able to use an EoS provided as a numerical table we need to approximate it in a way that is consistent with the previously discussed perturbed TOV equations. To this end we use the following methodology
\begin{itemize}
    \item First, we discretize the EoS $(p,\epsilon)$ parametric space into equally spaced steps.
    \item We fit a polytrope into every one of these steps and run tests with several choices of discretization steps to recover the option that best fits the CompOSE data for the SLy4 EoS. We have checked that, as expected, a smaller step produces a better fit.
    \item Our numerical EoS (pressure and energy density data) is transformed into a $(n_i,K_i)$ table, where the $n_i,K_i$ are the polytropic index and polytropic coefficient for step $i$ respectively. As a result, we have approximated our EoS using piecewise polytopic parameters, which can be used to solve the TOV equations. It is important to note that the $(n_i,K_i)$ are merely phenomenological parameters, that is they do not hold any physical meaning about the properties of matter. As a result, $n$ values outside the relativistic and non-relativistic limits are in principle allowed.
    \item In previous sections, we used built-in Python solvers for the TOV problem. But now, since the polytrope changes for each step, as we move from the center of the star outwards, we require an algorithm that allows for changing numerical integration parameter values with every step. As a result, we developed an algorithm from first principles based on the 4th-order Runge-Kutta method to have full control over the equations that are integrated. 
\end{itemize}
These steps enable the extension of our exploration of the white dwarf mass-radius relation, reported in the previous section, to neutron stars with numerical EoS.

\begin{figure*}
  \begin{center}
    \includegraphics[width=0.66\columnwidth]{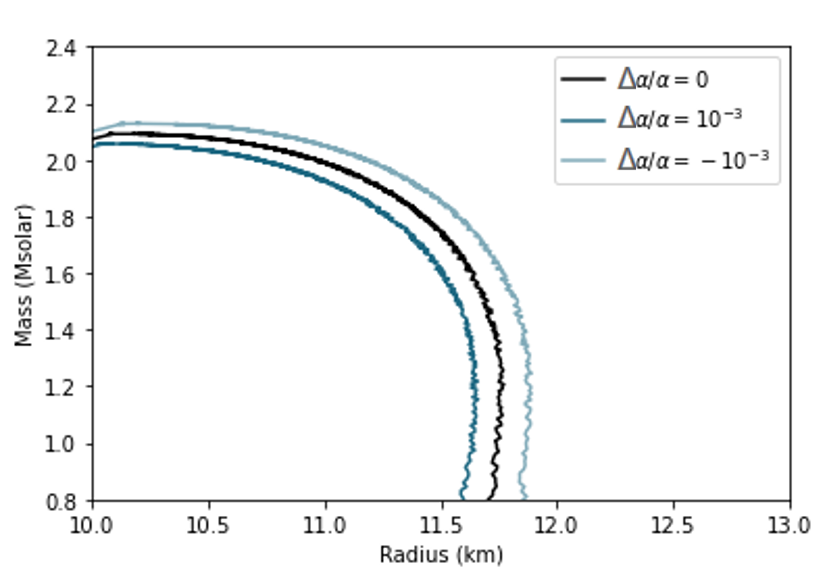}
    \includegraphics[width=0.66\columnwidth]{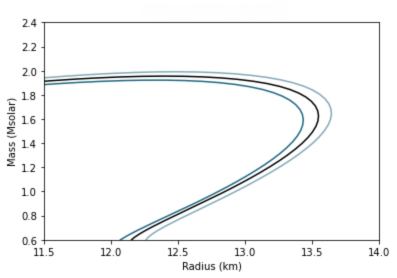}
    \includegraphics[width=0.66\columnwidth]{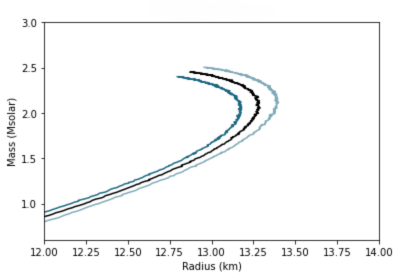}
    \includegraphics[width=0.66\columnwidth]{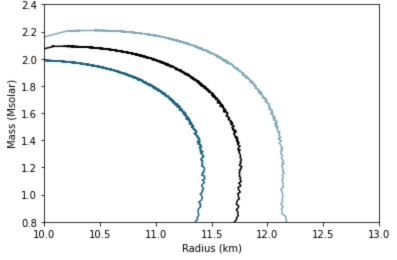}
    \includegraphics[width=0.66\columnwidth]{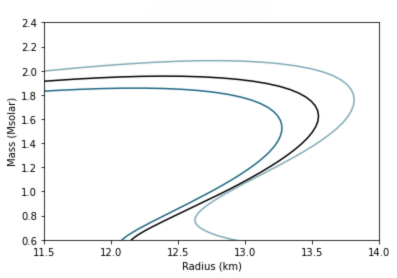}
    \includegraphics[width=0.66\columnwidth]{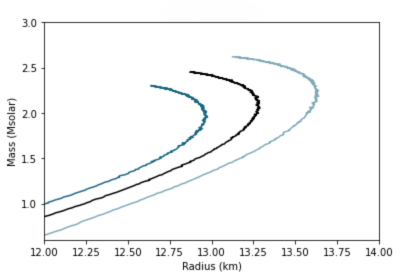}
    \includegraphics[width=0.66\columnwidth]{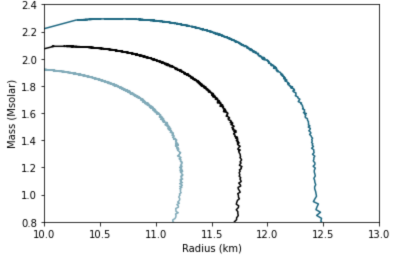}
    \includegraphics[width=0.66\columnwidth]{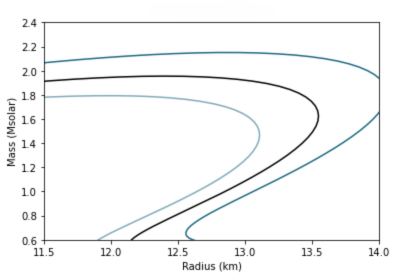}
    \includegraphics[width=0.66\columnwidth]{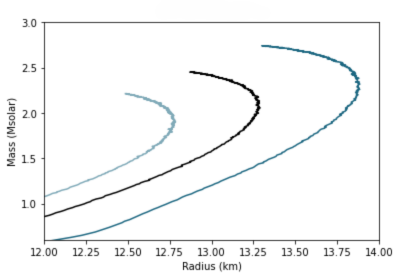}
    \caption{Neutron star Mass-Radius relation for the Unification, Dilaton and UV cutoff models (top, middle, and bottom panels respectively) for the three choices of EoS: SLy4 (left), FSU2R (middle) and DD2 (right). Positive and negative variations of $\alpha$ are depicted in dark and light blue respectively.}
    \label{fig3}
  \end{center}
\end{figure*}

We follow the same procedure as in the white dwarf model, investigating 3 different GUT cases. We use the above methodology to produce results for numerical EoS, assuming positive and negative perturbations of $\alpha$ of the order of $10^{-3}$.

Figure \ref{fig3} shows the results for all combinations of the three models and three EoS. Starting with the Unification case. we again find that negative perturbations raise the mass and radius of the star, whereas positive variations have the opposite effect. It is reasonable to recover similar results to the polytropic white dwarf since neutron stars rely on the gravity-degeneracy pressure equilibrium for stability. Consequently, the GUT model and $\alpha$ variation will affect the QCD scale and neutron degeneracy pressure respectively, leading to a shift in the equilibrium and the Mass-Radius relation. In the Dilaton and UV-Cutoff models, we observe larger deviations from the unperturbed case for the same amount of $\alpha$ variation, for the reasons already addressed in the previous section. We also confirm that the effects of positive and negative $\alpha$ variations are reversed for the UV cutoff model with respect to the others, due to the negative $R$ and relatively small $S$ parameters therein. Another expected result is the asymmetry of the perturbed cases, which is more noticeable in the UV cutoff case but exists in all due to the additional perturbation in the first relativistic correction.

While the study of EoS and GUT presented here is not exhaustive, given the large number or alternatives that exist, the solutions discussed provide significant insight on the possible combinations that are permitted once combined with neutron star mass measurements. The most massive neutron stars observed have the following masses: $2.08 \pm0.07~M_{\odot}$ for pulsar J0740+6620 \cite{2021ApJ...915L..12F} and $2.01\pm 0.04~M_{\odot}$ for pulsar J0348+0432 \cite{2013Sci...340..448A}. These measurements can rule out EoS whose maximum masses are below these values. Quite remarkably though, some EoS, such as the FSU2R that marginally reaches the $2M_{\odot}$ limit, have maximum masses that clearly exceed this limit under a perturbation of $\Delta \alpha /\alpha =-10^{-3}$ and Dilaton and UV-Cutoff models, thus making them likely candidates for the equation of state of dense matter. Similarly, the tendency of positive values $\Delta \alpha /\alpha$ to push towards smaller maximum masses, suggests that such combinations are unlikely unless EoS with very high maximum masses are considered. A thorough exploration of different EoS and perturbations of GUT in the light of recent measurements of neutron star masses and radii, using a statistical approach, is worthy and reserved for future study. 

\section{\label{sec:concl}Conclusions}

Our work addressed the impact of the variation of fundamental couplings on compact objects. We assumed a broad class of GUT models in which dilaton-type scalar fields driving $\alpha$ variations, and specifically considered three representative examples of GUT models, describing details of these expected variations. We explored the potential for constraining this model class using observations of white dwarfs and neutron stars.

For polytropic white dwarfs, we followed a perturbative approach, modeling the impact of $\alpha$ variations as small deviations from the unperturbed Mass-Radius relation. Results show significant (potentially observable) deviations from the unperturbed case for relative variations at the level of $\frac{\Delta\alpha}{\alpha}\sim10^{-3}$ in the Unification Model, which gets more pronounced with increasing $R$  (which is the case of the Dilaton Model) for reasonable $S$ values. The UV-Cutoff Model presents a reversed behavior, due to the negative $R$ dominating the leading order perturbation term. Plots of the Chandrasekhar mass in Fig.\ref{fig2} show the  GUT model-dependent nature of constraints on the $\alpha$ variation.

We extended this methodology to neutron stars by interpolating/extrapolating the white dwarf analytical solutions, while also imposing numerical EoS choices and adopting less confining computational methods. Neutron star results agree with the ones acquired from the white dwarf model, proving the robustness of our methodology. Here the specific choice of EoS is an additional source of degeneracy (in the statistical sense of the word) between various choices of model parameters. While we believe that our choices of GUT models and EoS are representative of the broad range of theoretical possibilities, they are certainly not exhaustive and a mode extensive exploration is warranted, especially when it comes to the EoS. Nevertheless, our analysis already shows that observational determinations of the range of neutron star masses are a promising route for joint constraints on the EoS and GUT-based models.

In addition to the insights gained from compact astrophysical objects, future work could benefit greatly from combining Mass-Radius observations with atomic clock measurements of the fine-structure constant. These independent yet complementary methods could provide priors on the $(R,S)$ parameter space, breaking some of the degeneracies among model parameters and thereby enhancing the robustness of our constraints. Overall, these studies serve the purpose of deepening our understanding $\alpha$'s role in fundamental physics paradigms, and of possible extensions of the Standard Model.

\section*{Acknowledgements}

This work was financed by Portuguese funds through FCT (Funda\c c\~ao para a Ci\^encia e a Tecnologia) in the framework of the project 2022.04048.PTDC (Phi in the Sky, DOI 10.54499/2022.04048.PTDC). CJM also acknowledges FCT and POCH/FSE (EC) support through Investigador FCT Contract 2021.01214.CEECIND/CP1658/CT0001 (DOI 10.54499/2021.01214.CEECIND/CP1658/CT0001).

This work was supported by computational time granted by the National Infrastructures for Research and Technology S.A. (GRNET S.A.) in the National HPC facility - ARIS - under project ID
pr017008/simnstar2.

\bibliography{article}

\providecommand{\noopsort}[1]{}\providecommand{\singleletter}[1]{#1}%
\begin{thebibliography}{48}%
\makeatletter
\providecommand \@ifxundefined [1]{%
 \@ifx{#1\undefined}
}%
\providecommand \@ifnum [1]{%
 \ifnum #1\expandafter \@firstoftwo
 \else \expandafter \@secondoftwo
 \fi
}%
\providecommand \@ifx [1]{%
 \ifx #1\expandafter \@firstoftwo
 \else \expandafter \@secondoftwo
 \fi
}%
\providecommand \natexlab [1]{#1}%
\providecommand \enquote  [1]{``#1''}%
\providecommand \bibnamefont  [1]{#1}%
\providecommand \bibfnamefont [1]{#1}%
\providecommand \citenamefont [1]{#1}%
\providecommand \href@noop [0]{\@secondoftwo}%
\providecommand \href [0]{\begingroup \@sanitize@url \@href}%
\providecommand \@href[1]{\@@startlink{#1}\@@href}%
\providecommand \@@href[1]{\endgroup#1\@@endlink}%
\providecommand \@sanitize@url [0]{\catcode `\\12\catcode `\$12\catcode
  `\&12\catcode `\#12\catcode `\^12\catcode `\_12\catcode `\%12\relax}%
\providecommand \@@startlink[1]{}%
\providecommand \@@endlink[0]{}%
\providecommand \url  [0]{\begingroup\@sanitize@url \@url }%
\providecommand \@url [1]{\endgroup\@href {#1}{\urlprefix }}%
\providecommand \urlprefix  [0]{URL }%
\providecommand \Eprint [0]{\href }%
\providecommand \doibase [0]{http://dx.doi.org/}%
\providecommand \selectlanguage [0]{\@gobble}%
\providecommand \bibinfo  [0]{\@secondoftwo}%
\providecommand \bibfield  [0]{\@secondoftwo}%
\providecommand \translation [1]{[#1]}%
\providecommand \BibitemOpen [0]{}%
\providecommand \bibitemStop [0]{}%
\providecommand \bibitemNoStop [0]{.\EOS\space}%
\providecommand \EOS [0]{\spacefactor3000\relax}%
\providecommand \BibitemShut  [1]{\csname bibitem#1\endcsname}%
\let\auto@bib@innerbib\@empty
\bibitem [{\citenamefont {Langacker}(1981)}]{Langacker}%
  \BibitemOpen
  \bibfield  {author} {\bibinfo {author} {\bibfnamefont {P.}~\bibnamefont
  {Langacker}},\ }\href {\doibase 10.1016/0370-1573(81)90059-4} {\bibfield
  {journal} {\bibinfo  {journal} {Phys. Rept.}\ }\textbf {\bibinfo {volume}
  {72}},\ \bibinfo {pages} {185} (\bibinfo {year} {1981})}\BibitemShut
  {NoStop}%
\bibitem [{\citenamefont {Navas}\ \emph {et~al.}(2024)\citenamefont {Navas}
  \emph {et~al.}}]{PDG}%
  \BibitemOpen
  \bibfield  {author} {\bibinfo {author} {\bibfnamefont {S.}~\bibnamefont
  {Navas}} \emph {et~al.} (\bibinfo {collaboration} {Particle Data Group}),\
  }\href {\doibase 10.1103/PhysRevD.110.030001} {\bibfield  {journal} {\bibinfo
   {journal} {Phys. Rev. D}\ }\textbf {\bibinfo {volume} {110}},\ \bibinfo
  {pages} {030001} (\bibinfo {year} {2024})}\BibitemShut {NoStop}%
\bibitem [{\citenamefont {Martins}(2017)}]{ROPP}%
  \BibitemOpen
  \bibfield  {author} {\bibinfo {author} {\bibfnamefont {C.~J. A.~P.}\
  \bibnamefont {Martins}},\ }\href {\doibase 10.1088/1361-6633/aa860e}
  {\bibfield  {journal} {\bibinfo  {journal} {Rep. Prog. Phys.}\ }\textbf
  {\bibinfo {volume} {80}},\ \bibinfo {pages} {126902} (\bibinfo {year}
  {2017})},\ \Eprint {http://arxiv.org/abs/1709.02923} {arXiv:1709.02923
  [astro-ph.CO]} \BibitemShut {NoStop}%
\bibitem [{\citenamefont {Uzan}(2024)}]{Uzan}%
  \BibitemOpen
  \bibfield  {author} {\bibinfo {author} {\bibfnamefont {J.-P.}\ \bibnamefont
  {Uzan}},\ }\href@noop {} {\  (\bibinfo {year} {2024})},\ \Eprint
  {http://arxiv.org/abs/2410.07281} {arXiv:2410.07281 [astro-ph.CO]}
  \BibitemShut {NoStop}%
\bibitem [{\citenamefont {Shapiro}\ and\ \citenamefont
  {Teukolsky}(1983)}]{Shapiro}%
  \BibitemOpen
  \bibfield  {author} {\bibinfo {author} {\bibfnamefont {S.~L.}\ \bibnamefont
  {Shapiro}}\ and\ \bibinfo {author} {\bibfnamefont {S.~A.}\ \bibnamefont
  {Teukolsky}},\ }\href {\doibase 10.1002/9783527617661} {\emph {\bibinfo
  {title} {{Black holes, white dwarfs, and neutron stars: The physics of
  compact objects}}}}\ (\bibinfo {year} {1983})\BibitemShut {NoStop}%
\bibitem [{\citenamefont {\"Ozel}\ and\ \citenamefont {Freire}(2016)}]{Ozel}%
  \BibitemOpen
  \bibfield  {author} {\bibinfo {author} {\bibfnamefont {F.}~\bibnamefont
  {\"Ozel}}\ and\ \bibinfo {author} {\bibfnamefont {P.}~\bibnamefont
  {Freire}},\ }\href {\doibase 10.1146/annurev-astro-081915-023322} {\bibfield
  {journal} {\bibinfo  {journal} {Ann. Rev. Astron. Astrophys.}\ }\textbf
  {\bibinfo {volume} {54}},\ \bibinfo {pages} {401} (\bibinfo {year} {2016})},\
  \Eprint {http://arxiv.org/abs/1603.02698} {arXiv:1603.02698 [astro-ph.HE]}
  \BibitemShut {NoStop}%
\bibitem [{\citenamefont {Berengut}\ \emph {et~al.}(2013)\citenamefont
  {Berengut}, \citenamefont {Flambaum}, \citenamefont {Ong}, \citenamefont
  {Webb}, \citenamefont {Barrow}, \citenamefont {Barstow}, \citenamefont
  {Preval},\ and\ \citenamefont {Holberg}}]{Berengut}%
  \BibitemOpen
  \bibfield  {author} {\bibinfo {author} {\bibfnamefont {J.~C.}\ \bibnamefont
  {Berengut}}, \bibinfo {author} {\bibfnamefont {V.~V.}\ \bibnamefont
  {Flambaum}}, \bibinfo {author} {\bibfnamefont {A.}~\bibnamefont {Ong}},
  \bibinfo {author} {\bibfnamefont {J.~K.}\ \bibnamefont {Webb}}, \bibinfo
  {author} {\bibfnamefont {J.~D.}\ \bibnamefont {Barrow}}, \bibinfo {author}
  {\bibfnamefont {M.~A.}\ \bibnamefont {Barstow}}, \bibinfo {author}
  {\bibfnamefont {S.~P.}\ \bibnamefont {Preval}}, \ and\ \bibinfo {author}
  {\bibfnamefont {J.~B.}\ \bibnamefont {Holberg}},\ }\href {\doibase
  10.1103/PhysRevLett.111.010801} {\bibfield  {journal} {\bibinfo  {journal}
  {Phys. Rev. Lett.}\ }\textbf {\bibinfo {volume} {111}},\ \bibinfo {pages}
  {010801} (\bibinfo {year} {2013})},\ \Eprint {http://arxiv.org/abs/1305.1337}
  {arXiv:1305.1337 [astro-ph.CO]} \BibitemShut {NoStop}%
\bibitem [{\citenamefont {Bagdonaite}\ \emph {et~al.}(2014)\citenamefont
  {Bagdonaite}, \citenamefont {Salumbides}, \citenamefont {Preval},
  \citenamefont {Barstow}, \citenamefont {Barrow}, \citenamefont {Murphy},\
  and\ \citenamefont {Ubachs}}]{Bagdonaite}%
  \BibitemOpen
  \bibfield  {author} {\bibinfo {author} {\bibfnamefont {J.}~\bibnamefont
  {Bagdonaite}}, \bibinfo {author} {\bibfnamefont {E.~J.}\ \bibnamefont
  {Salumbides}}, \bibinfo {author} {\bibfnamefont {S.~P.}\ \bibnamefont
  {Preval}}, \bibinfo {author} {\bibfnamefont {M.~A.}\ \bibnamefont {Barstow}},
  \bibinfo {author} {\bibfnamefont {J.~D.}\ \bibnamefont {Barrow}}, \bibinfo
  {author} {\bibfnamefont {M.~T.}\ \bibnamefont {Murphy}}, \ and\ \bibinfo
  {author} {\bibfnamefont {W.}~\bibnamefont {Ubachs}},\ }\href {\doibase
  10.1103/PhysRevLett.113.123002} {\bibfield  {journal} {\bibinfo  {journal}
  {Phys. Rev. Lett.}\ }\textbf {\bibinfo {volume} {113}},\ \bibinfo {pages}
  {123002} (\bibinfo {year} {2014})},\ \Eprint {http://arxiv.org/abs/1409.1000}
  {arXiv:1409.1000 [gr-qc]} \BibitemShut {NoStop}%
\bibitem [{\citenamefont {Olive}\ and\ \citenamefont
  {Pospelov}(2008{\natexlab{a}})}]{Olive}%
  \BibitemOpen
  \bibfield  {author} {\bibinfo {author} {\bibfnamefont {K.~A.}\ \bibnamefont
  {Olive}}\ and\ \bibinfo {author} {\bibfnamefont {M.}~\bibnamefont
  {Pospelov}},\ }\href {\doibase 10.1103/PhysRevD.77.043524} {\bibfield
  {journal} {\bibinfo  {journal} {Phys. Rev. D}\ }\textbf {\bibinfo {volume}
  {77}},\ \bibinfo {pages} {043524} (\bibinfo {year} {2008}{\natexlab{a}})},\
  \Eprint {http://arxiv.org/abs/0709.3825} {arXiv:0709.3825 [hep-ph]}
  \BibitemShut {NoStop}%
\bibitem [{\citenamefont {Jain}\ \emph {et~al.}(2016)\citenamefont {Jain},
  \citenamefont {Kouvaris},\ and\ \citenamefont {Nielsen}}]{Jain}%
  \BibitemOpen
  \bibfield  {author} {\bibinfo {author} {\bibfnamefont {R.~K.}\ \bibnamefont
  {Jain}}, \bibinfo {author} {\bibfnamefont {C.}~\bibnamefont {Kouvaris}}, \
  and\ \bibinfo {author} {\bibfnamefont {N.~G.}\ \bibnamefont {Nielsen}},\
  }\href {\doibase 10.1103/PhysRevLett.116.151103} {\bibfield  {journal}
  {\bibinfo  {journal} {Phys. Rev. Lett.}\ }\textbf {\bibinfo {volume} {116}},\
  \bibinfo {pages} {151103} (\bibinfo {year} {2016})},\ \Eprint
  {http://arxiv.org/abs/1512.05946} {arXiv:1512.05946 [astro-ph.CO]}
  \BibitemShut {NoStop}%
\bibitem [{\citenamefont {Magano}\ \emph {et~al.}(2017)\citenamefont {Magano},
  \citenamefont {Vilas~Boas},\ and\ \citenamefont {Martins}}]{Magano}%
  \BibitemOpen
  \bibfield  {author} {\bibinfo {author} {\bibfnamefont {D.~M.~N.}\
  \bibnamefont {Magano}}, \bibinfo {author} {\bibfnamefont {J.~M.~A.}\
  \bibnamefont {Vilas~Boas}}, \ and\ \bibinfo {author} {\bibfnamefont {C.~J.
  A.~P.}\ \bibnamefont {Martins}},\ }\href {\doibase
  10.1103/PhysRevD.96.083012} {\bibfield  {journal} {\bibinfo  {journal} {Phys.
  Rev. D}\ }\textbf {\bibinfo {volume} {96}},\ \bibinfo {pages} {083012}
  (\bibinfo {year} {2017})},\ \Eprint {http://arxiv.org/abs/1710.05828}
  {arXiv:1710.05828 [astro-ph.CO]} \BibitemShut {NoStop}%
\bibitem [{\citenamefont {Aad}\ \emph {et~al.}(2012)\citenamefont {Aad} \emph
  {et~al.}}]{ATLAS}%
  \BibitemOpen
  \bibfield  {author} {\bibinfo {author} {\bibfnamefont {G.}~\bibnamefont
  {Aad}} \emph {et~al.} (\bibinfo {collaboration} {ATLAS}),\ }\href {\doibase
  10.1016/j.physletb.2012.08.020} {\bibfield  {journal} {\bibinfo  {journal}
  {Phys. Lett. B}\ }\textbf {\bibinfo {volume} {716}},\ \bibinfo {pages} {1}
  (\bibinfo {year} {2012})},\ \Eprint {http://arxiv.org/abs/1207.7214}
  {arXiv:1207.7214 [hep-ex]} \BibitemShut {NoStop}%
\bibitem [{\citenamefont {Chatrchyan}\ \emph {et~al.}(2012)\citenamefont
  {Chatrchyan} \emph {et~al.}}]{CMS}%
  \BibitemOpen
  \bibfield  {author} {\bibinfo {author} {\bibfnamefont {S.}~\bibnamefont
  {Chatrchyan}} \emph {et~al.} (\bibinfo {collaboration} {CMS}),\ }\href
  {\doibase 10.1016/j.physletb.2012.08.021} {\bibfield  {journal} {\bibinfo
  {journal} {Phys. Lett. B}\ }\textbf {\bibinfo {volume} {716}},\ \bibinfo
  {pages} {30} (\bibinfo {year} {2012})},\ \Eprint
  {http://arxiv.org/abs/1207.7235} {arXiv:1207.7235 [hep-ex]} \BibitemShut
  {NoStop}%
\bibitem [{\citenamefont {{Carroll}}(1998)}]{1998PhRvL..81.3067C}%
  \BibitemOpen
  \bibfield  {author} {\bibinfo {author} {\bibfnamefont {S.~M.}\ \bibnamefont
  {{Carroll}}},\ }\href {\doibase 10.1103/PhysRevLett.81.3067} {\bibfield
  {journal} {\bibinfo  {journal} {\prl}\ }\textbf {\bibinfo {volume} {81}},\
  \bibinfo {pages} {3067} (\bibinfo {year} {1998})},\ \Eprint
  {http://arxiv.org/abs/astro-ph/9806099} {arXiv:astro-ph/9806099 [astro-ph]}
  \BibitemShut {NoStop}%
\bibitem [{\citenamefont {Olive}\ and\ \citenamefont
  {Pospelov}(2008{\natexlab{b}})}]{OlivePospelov}%
  \BibitemOpen
  \bibfield  {author} {\bibinfo {author} {\bibfnamefont {K.~A.}\ \bibnamefont
  {Olive}}\ and\ \bibinfo {author} {\bibfnamefont {M.}~\bibnamefont
  {Pospelov}},\ }\href {\doibase 10.1103/PhysRevD.77.043524} {\bibfield
  {journal} {\bibinfo  {journal} {Phys. Rev. D}\ }\textbf {\bibinfo {volume}
  {77}},\ \bibinfo {pages} {043524} (\bibinfo {year} {2008}{\natexlab{b}})},\
  \Eprint {http://arxiv.org/abs/0709.3825} {arXiv:0709.3825 [hep-ph]}
  \BibitemShut {NoStop}%
\bibitem [{\citenamefont {Coc}\ \emph {et~al.}(2007)\citenamefont {Coc},
  \citenamefont {Nunes}, \citenamefont {Olive}, \citenamefont {Uzan},\ and\
  \citenamefont {Vangioni}}]{Coc}%
  \BibitemOpen
  \bibfield  {author} {\bibinfo {author} {\bibfnamefont {A.}~\bibnamefont
  {Coc}}, \bibinfo {author} {\bibfnamefont {N.~J.}\ \bibnamefont {Nunes}},
  \bibinfo {author} {\bibfnamefont {K.~A.}\ \bibnamefont {Olive}}, \bibinfo
  {author} {\bibfnamefont {J.-P.}\ \bibnamefont {Uzan}}, \ and\ \bibinfo
  {author} {\bibfnamefont {E.}~\bibnamefont {Vangioni}},\ }\href {\doibase
  10.1103/PhysRevD.76.023511} {\bibfield  {journal} {\bibinfo  {journal} {Phys.
  Rev. D}\ }\textbf {\bibinfo {volume} {76}},\ \bibinfo {pages} {023511}
  (\bibinfo {year} {2007})},\ \Eprint {http://arxiv.org/abs/astro-ph/0610733}
  {arXiv:astro-ph/0610733} \BibitemShut {NoStop}%
\bibitem [{\citenamefont {Campbell}\ and\ \citenamefont
  {Olive}(1995)}]{Campbell}%
  \BibitemOpen
  \bibfield  {author} {\bibinfo {author} {\bibfnamefont {B.~A.}\ \bibnamefont
  {Campbell}}\ and\ \bibinfo {author} {\bibfnamefont {K.~A.}\ \bibnamefont
  {Olive}},\ }\href {\doibase 10.1016/0370-2693(94)01652-S} {\bibfield
  {journal} {\bibinfo  {journal} {Phys. Lett. B}\ }\textbf {\bibinfo {volume}
  {345}},\ \bibinfo {pages} {429} (\bibinfo {year} {1995})},\ \Eprint
  {http://arxiv.org/abs/hep-ph/9411272} {arXiv:hep-ph/9411272} \BibitemShut
  {NoStop}%
\bibitem [{\citenamefont {Vieira}\ \emph {et~al.}(2012)\citenamefont {Vieira},
  \citenamefont {Martins},\ and\ \citenamefont {Monteiro}}]{Vieira}%
  \BibitemOpen
  \bibfield  {author} {\bibinfo {author} {\bibfnamefont {J.~P.~P.}\
  \bibnamefont {Vieira}}, \bibinfo {author} {\bibfnamefont {C.~J. A.~P.}\
  \bibnamefont {Martins}}, \ and\ \bibinfo {author} {\bibfnamefont {M.~J. P.
  F.~G.}\ \bibnamefont {Monteiro}},\ }\href {\doibase
  10.1103/PhysRevD.86.043003} {\bibfield  {journal} {\bibinfo  {journal} {Phys.
  Rev. D}\ }\textbf {\bibinfo {volume} {86}},\ \bibinfo {pages} {043003}
  (\bibinfo {year} {2012})},\ \Eprint {http://arxiv.org/abs/1206.3665}
  {arXiv:1206.3665 [astro-ph.SR]} \BibitemShut {NoStop}%
\bibitem [{\citenamefont {Ekstrom}\ \emph {et~al.}(2010)\citenamefont
  {Ekstrom}, \citenamefont {Coc}, \citenamefont {Descouvemont}, \citenamefont
  {Meynet}, \citenamefont {Olive}, \citenamefont {Uzan},\ and\ \citenamefont
  {Vangioni}}]{Ekstrom}%
  \BibitemOpen
  \bibfield  {author} {\bibinfo {author} {\bibfnamefont {S.}~\bibnamefont
  {Ekstrom}}, \bibinfo {author} {\bibfnamefont {A.}~\bibnamefont {Coc}},
  \bibinfo {author} {\bibfnamefont {P.}~\bibnamefont {Descouvemont}}, \bibinfo
  {author} {\bibfnamefont {G.}~\bibnamefont {Meynet}}, \bibinfo {author}
  {\bibfnamefont {K.~A.}\ \bibnamefont {Olive}}, \bibinfo {author}
  {\bibfnamefont {J.-P.}\ \bibnamefont {Uzan}}, \ and\ \bibinfo {author}
  {\bibfnamefont {E.}~\bibnamefont {Vangioni}},\ }\href {\doibase
  10.1051/0004-6361/200913684} {\bibfield  {journal} {\bibinfo  {journal}
  {Astron. Astrophys.}\ }\textbf {\bibinfo {volume} {514}},\ \bibinfo {pages}
  {A62} (\bibinfo {year} {2010})},\ \Eprint {http://arxiv.org/abs/0911.2420}
  {arXiv:0911.2420 [astro-ph.SR]} \BibitemShut {NoStop}%
\bibitem [{\citenamefont {Perez-Garcia}\ and\ \citenamefont
  {Martins}(2012)}]{Perez}%
  \BibitemOpen
  \bibfield  {author} {\bibinfo {author} {\bibfnamefont {M.~A.}\ \bibnamefont
  {Perez-Garcia}}\ and\ \bibinfo {author} {\bibfnamefont {C.~J. A.~P.}\
  \bibnamefont {Martins}},\ }\href {\doibase 10.1016/j.physletb.2012.10.047}
  {\bibfield  {journal} {\bibinfo  {journal} {Phys. Lett. B}\ }\textbf
  {\bibinfo {volume} {718}},\ \bibinfo {pages} {241} (\bibinfo {year}
  {2012})},\ \Eprint {http://arxiv.org/abs/1203.0399} {arXiv:1203.0399
  [astro-ph.CO]} \BibitemShut {NoStop}%
\bibitem [{\citenamefont {Gasser}\ and\ \citenamefont
  {Leutwyler}(1982)}]{Gasser}%
  \BibitemOpen
  \bibfield  {author} {\bibinfo {author} {\bibfnamefont {J.}~\bibnamefont
  {Gasser}}\ and\ \bibinfo {author} {\bibfnamefont {H.}~\bibnamefont
  {Leutwyler}},\ }\href {\doibase 10.1016/0370-1573(82)90035-7} {\bibfield
  {journal} {\bibinfo  {journal} {Phys. Rept.}\ }\textbf {\bibinfo {volume}
  {87}},\ \bibinfo {pages} {77} (\bibinfo {year} {1982})}\BibitemShut {NoStop}%
\bibitem [{\citenamefont {Nakashima}\ \emph {et~al.}(2010)\citenamefont
  {Nakashima}, \citenamefont {Ichikawa}, \citenamefont {Nagata},\ and\
  \citenamefont {Yokoyama}}]{Nakashima}%
  \BibitemOpen
  \bibfield  {author} {\bibinfo {author} {\bibfnamefont {M.}~\bibnamefont
  {Nakashima}}, \bibinfo {author} {\bibfnamefont {K.}~\bibnamefont {Ichikawa}},
  \bibinfo {author} {\bibfnamefont {R.}~\bibnamefont {Nagata}}, \ and\ \bibinfo
  {author} {\bibfnamefont {J.}~\bibnamefont {Yokoyama}},\ }\href {\doibase
  10.1088/1475-7516/2010/01/030} {\bibfield  {journal} {\bibinfo  {journal}
  {JCAP}\ }\textbf {\bibinfo {volume} {01}},\ \bibinfo {pages} {030} (\bibinfo
  {year} {2010})},\ \Eprint {http://arxiv.org/abs/0910.0742} {arXiv:0910.0742
  [astro-ph.CO]} \BibitemShut {NoStop}%
\bibitem [{\citenamefont {Lee}(2024)}]{Lee}%
  \BibitemOpen
  \bibfield  {author} {\bibinfo {author} {\bibfnamefont {T.}~\bibnamefont
  {Lee}},\ }\href {\doibase 10.1016/j.physletb.2023.138424} {\bibfield
  {journal} {\bibinfo  {journal} {Phys. Lett. B}\ }\textbf {\bibinfo {volume}
  {849}},\ \bibinfo {pages} {138424} (\bibinfo {year} {2024})},\ \Eprint
  {http://arxiv.org/abs/2310.13308} {arXiv:2310.13308 [hep-ph]} \BibitemShut
  {NoStop}%
\bibitem [{\citenamefont {{Heger}}\ \emph {et~al.}(2003)\citenamefont
  {{Heger}}, \citenamefont {{Fryer}}, \citenamefont {{Woosley}}, \citenamefont
  {{Langer}},\ and\ \citenamefont {{Hartmann}}}]{2003ApJ...591..288H}%
  \BibitemOpen
  \bibfield  {author} {\bibinfo {author} {\bibfnamefont {A.}~\bibnamefont
  {{Heger}}}, \bibinfo {author} {\bibfnamefont {C.~L.}\ \bibnamefont
  {{Fryer}}}, \bibinfo {author} {\bibfnamefont {S.~E.}\ \bibnamefont
  {{Woosley}}}, \bibinfo {author} {\bibfnamefont {N.}~\bibnamefont {{Langer}}},
  \ and\ \bibinfo {author} {\bibfnamefont {D.~H.}\ \bibnamefont {{Hartmann}}},\
  }\href {\doibase 10.1086/375341} {\bibfield  {journal} {\bibinfo  {journal}
  {\apj}\ }\textbf {\bibinfo {volume} {591}},\ \bibinfo {pages} {288} (\bibinfo
  {year} {2003})},\ \Eprint {http://arxiv.org/abs/astro-ph/0212469}
  {arXiv:astro-ph/0212469 [astro-ph]} \BibitemShut {NoStop}%
\bibitem [{\citenamefont {Silbar}\ and\ \citenamefont {Reddy}(2004)}]{Silbar}%
  \BibitemOpen
  \bibfield  {author} {\bibinfo {author} {\bibfnamefont {R.~R.}\ \bibnamefont
  {Silbar}}\ and\ \bibinfo {author} {\bibfnamefont {S.}~\bibnamefont {Reddy}},\
  }\href {\doibase 10.1119/1.1852544} {\bibfield  {journal} {\bibinfo
  {journal} {Am. J. Phys.}\ }\textbf {\bibinfo {volume} {72}},\ \bibinfo
  {pages} {892} (\bibinfo {year} {2004})},\ \bibinfo {note} {[Erratum:
  Am.J.Phys. 73, 286 (2005)]},\ \Eprint {http://arxiv.org/abs/nucl-th/0309041}
  {arXiv:nucl-th/0309041} \BibitemShut {NoStop}%
\bibitem [{\citenamefont {Sagert}\ \emph {et~al.}(2006)\citenamefont {Sagert},
  \citenamefont {Hempel}, \citenamefont {Greiner},\ and\ \citenamefont
  {Schaffner-Bielich}}]{Sagert}%
  \BibitemOpen
  \bibfield  {author} {\bibinfo {author} {\bibfnamefont {I.}~\bibnamefont
  {Sagert}}, \bibinfo {author} {\bibfnamefont {M.}~\bibnamefont {Hempel}},
  \bibinfo {author} {\bibfnamefont {C.}~\bibnamefont {Greiner}}, \ and\
  \bibinfo {author} {\bibfnamefont {J.}~\bibnamefont {Schaffner-Bielich}},\
  }\href {\doibase 10.1088/0143-0807/27/3/012} {\bibfield  {journal} {\bibinfo
  {journal} {Eur. J. Phys.}\ }\textbf {\bibinfo {volume} {27}},\ \bibinfo
  {pages} {577} (\bibinfo {year} {2006})},\ \Eprint
  {http://arxiv.org/abs/astro-ph/0506417} {arXiv:astro-ph/0506417} \BibitemShut
  {NoStop}%
\bibitem [{\citenamefont {{Chandrasekhar}}(1939)}]{Chand}%
  \BibitemOpen
  \bibfield  {author} {\bibinfo {author} {\bibfnamefont {S.}~\bibnamefont
  {{Chandrasekhar}}},\ }\href@noop {} {\emph {\bibinfo {title} {{An
  introduction to the study of stellar structure}}}}\ (\bibinfo {year}
  {1939})\BibitemShut {NoStop}%
\bibitem [{\citenamefont {{Caiazzo}}\ \emph {et~al.}(2021)\citenamefont
  {{Caiazzo}}, \citenamefont {{Burdge}}, \citenamefont {{Fuller}},
  \citenamefont {{Heyl}}, \citenamefont {{Kulkarni}}, \citenamefont {{Prince}},
  \citenamefont {{Richer}}, \citenamefont {{Schwab}}, \citenamefont
  {{Andreoni}}, \citenamefont {{Bellm}}, \citenamefont {{Drake}}, \citenamefont
  {{Duev}}, \citenamefont {{Graham}}, \citenamefont {{Helou}}, \citenamefont
  {{Mahabal}}, \citenamefont {{Masci}}, \citenamefont {{Smith}},\ and\
  \citenamefont {{Soumagnac}}}]{2021Natur.595...39C}%
  \BibitemOpen
  \bibfield  {author} {\bibinfo {author} {\bibfnamefont {I.}~\bibnamefont
  {{Caiazzo}}}, \bibinfo {author} {\bibfnamefont {K.~B.}\ \bibnamefont
  {{Burdge}}}, \bibinfo {author} {\bibfnamefont {J.}~\bibnamefont {{Fuller}}},
  \bibinfo {author} {\bibfnamefont {J.}~\bibnamefont {{Heyl}}}, \bibinfo
  {author} {\bibfnamefont {S.~R.}\ \bibnamefont {{Kulkarni}}}, \bibinfo
  {author} {\bibfnamefont {T.~A.}\ \bibnamefont {{Prince}}}, \bibinfo {author}
  {\bibfnamefont {H.~B.}\ \bibnamefont {{Richer}}}, \bibinfo {author}
  {\bibfnamefont {J.}~\bibnamefont {{Schwab}}}, \bibinfo {author}
  {\bibfnamefont {I.}~\bibnamefont {{Andreoni}}}, \bibinfo {author}
  {\bibfnamefont {E.~C.}\ \bibnamefont {{Bellm}}}, \bibinfo {author}
  {\bibfnamefont {A.}~\bibnamefont {{Drake}}}, \bibinfo {author} {\bibfnamefont
  {D.~A.}\ \bibnamefont {{Duev}}}, \bibinfo {author} {\bibfnamefont {M.~J.}\
  \bibnamefont {{Graham}}}, \bibinfo {author} {\bibfnamefont {G.}~\bibnamefont
  {{Helou}}}, \bibinfo {author} {\bibfnamefont {A.~A.}\ \bibnamefont
  {{Mahabal}}}, \bibinfo {author} {\bibfnamefont {F.~J.}\ \bibnamefont
  {{Masci}}}, \bibinfo {author} {\bibfnamefont {R.}~\bibnamefont {{Smith}}}, \
  and\ \bibinfo {author} {\bibfnamefont {M.~T.}\ \bibnamefont {{Soumagnac}}},\
  }\href {\doibase 10.1038/s41586-021-03615-y} {\bibfield  {journal} {\bibinfo
  {journal} {\nat}\ }\textbf {\bibinfo {volume} {595}},\ \bibinfo {pages} {39}
  (\bibinfo {year} {2021})},\ \Eprint {http://arxiv.org/abs/2107.08458}
  {arXiv:2107.08458 [astro-ph.SR]} \BibitemShut {NoStop}%
\bibitem [{\citenamefont {{Gulminelli}}\ and\ \citenamefont
  {{Raduta}}(2015)}]{2015PhRvC..92e5803G}%
  \BibitemOpen
  \bibfield  {author} {\bibinfo {author} {\bibfnamefont {F.}~\bibnamefont
  {{Gulminelli}}}\ and\ \bibinfo {author} {\bibfnamefont {A.~R.}\ \bibnamefont
  {{Raduta}}},\ }\href {\doibase 10.1103/PhysRevC.92.055803} {\bibfield
  {journal} {\bibinfo  {journal} {\prc}\ }\textbf {\bibinfo {volume} {92}},\
  \bibinfo {eid} {055803} (\bibinfo {year} {2015})},\ \Eprint
  {http://arxiv.org/abs/1504.04493} {arXiv:1504.04493 [nucl-th]} \BibitemShut
  {NoStop}%
\bibitem [{\citenamefont {Raduta}\ and\ \citenamefont
  {Gulminelli}(2019)}]{RADUTA2019252}%
  \BibitemOpen
  \bibfield  {author} {\bibinfo {author} {\bibfnamefont {A.}~\bibnamefont
  {Raduta}}\ and\ \bibinfo {author} {\bibfnamefont {F.}~\bibnamefont
  {Gulminelli}},\ }\href {\doibase
  https://doi.org/10.1016/j.nuclphysa.2018.11.003} {\bibfield  {journal}
  {\bibinfo  {journal} {Nuclear Physics A}\ }\textbf {\bibinfo {volume}
  {983}},\ \bibinfo {pages} {252} (\bibinfo {year} {2019})}\BibitemShut
  {NoStop}%
\bibitem [{\citenamefont {{Typel}}\ \emph {et~al.}(2015)\citenamefont
  {{Typel}}, \citenamefont {{Oertel}},\ and\ \citenamefont
  {{Kl{\"a}hn}}}]{2015PPN....46..633T}%
  \BibitemOpen
  \bibfield  {author} {\bibinfo {author} {\bibfnamefont {S.}~\bibnamefont
  {{Typel}}}, \bibinfo {author} {\bibfnamefont {M.}~\bibnamefont {{Oertel}}}, \
  and\ \bibinfo {author} {\bibfnamefont {T.}~\bibnamefont {{Kl{\"a}hn}}},\
  }\href {\doibase 10.1134/S1063779615040061} {\bibfield  {journal} {\bibinfo
  {journal} {Physics of Particles and Nuclei}\ }\textbf {\bibinfo {volume}
  {46}},\ \bibinfo {pages} {633} (\bibinfo {year} {2015})}\BibitemShut
  {NoStop}%
\bibitem [{\citenamefont {{Oertel}}\ \emph {et~al.}(2017)\citenamefont
  {{Oertel}}, \citenamefont {{Hempel}}, \citenamefont {{Kl{\"a}hn}},\ and\
  \citenamefont {{Typel}}}]{2017RvMP...89a5007O}%
  \BibitemOpen
  \bibfield  {author} {\bibinfo {author} {\bibfnamefont {M.}~\bibnamefont
  {{Oertel}}}, \bibinfo {author} {\bibfnamefont {M.}~\bibnamefont {{Hempel}}},
  \bibinfo {author} {\bibfnamefont {T.}~\bibnamefont {{Kl{\"a}hn}}}, \ and\
  \bibinfo {author} {\bibfnamefont {S.}~\bibnamefont {{Typel}}},\ }\href
  {\doibase 10.1103/RevModPhys.89.015007} {\bibfield  {journal} {\bibinfo
  {journal} {Reviews of Modern Physics}\ }\textbf {\bibinfo {volume} {89}},\
  \bibinfo {eid} {015007} (\bibinfo {year} {2017})},\ \Eprint
  {http://arxiv.org/abs/1610.03361} {arXiv:1610.03361 [astro-ph.HE]}
  \BibitemShut {NoStop}%
\bibitem [{\citenamefont {{CompOSE Core Team}}\ \emph
  {et~al.}(2022)\citenamefont {{CompOSE Core Team}}, \citenamefont {{Typel}},
  \citenamefont {{Oertel}}, \citenamefont {{Kl{\"a}hn}}, \citenamefont
  {{Chatterjee}}, \citenamefont {{Dexheimer}}, \citenamefont {{Ishizuka}},
  \citenamefont {{Mancini}}, \citenamefont {{Novak}}, \citenamefont {{Pais}},
  \citenamefont {{Provid{\^e}ncia}}, \citenamefont {{R. Raduta}}, \citenamefont
  {{Servillat}},\ and\ \citenamefont {{Tolos}}}]{2022EPJA...58..221C}%
  \BibitemOpen
  \bibfield  {author} {\bibinfo {author} {\bibnamefont {{CompOSE Core Team}}},
  \bibinfo {author} {\bibfnamefont {S.}~\bibnamefont {{Typel}}}, \bibinfo
  {author} {\bibfnamefont {M.}~\bibnamefont {{Oertel}}}, \bibinfo {author}
  {\bibfnamefont {T.}~\bibnamefont {{Kl{\"a}hn}}}, \bibinfo {author}
  {\bibfnamefont {D.}~\bibnamefont {{Chatterjee}}}, \bibinfo {author}
  {\bibfnamefont {V.}~\bibnamefont {{Dexheimer}}}, \bibinfo {author}
  {\bibfnamefont {C.}~\bibnamefont {{Ishizuka}}}, \bibinfo {author}
  {\bibfnamefont {M.}~\bibnamefont {{Mancini}}}, \bibinfo {author}
  {\bibfnamefont {J.}~\bibnamefont {{Novak}}}, \bibinfo {author} {\bibfnamefont
  {H.}~\bibnamefont {{Pais}}}, \bibinfo {author} {\bibfnamefont
  {C.}~\bibnamefont {{Provid{\^e}ncia}}}, \bibinfo {author} {\bibfnamefont
  {A.}~\bibnamefont {{R. Raduta}}}, \bibinfo {author} {\bibfnamefont
  {M.}~\bibnamefont {{Servillat}}}, \ and\ \bibinfo {author} {\bibfnamefont
  {L.}~\bibnamefont {{Tolos}}},\ }\href {\doibase
  10.1140/epja/s10050-022-00847-y} {\bibfield  {journal} {\bibinfo  {journal}
  {European Physical Journal A}\ }\textbf {\bibinfo {volume} {58}},\ \bibinfo
  {eid} {221} (\bibinfo {year} {2022})},\ \Eprint
  {http://arxiv.org/abs/2203.03209} {arXiv:2203.03209 [astro-ph.HE]}
  \BibitemShut {NoStop}%
\bibitem [{\citenamefont {Grill}\ \emph {et~al.}(2014)\citenamefont {Grill},
  \citenamefont {Pais}, \citenamefont {Provid\^encia}, \citenamefont
  {Vida\~na},\ and\ \citenamefont {Avancini}}]{PhysRevC.90.045803}%
  \BibitemOpen
  \bibfield  {author} {\bibinfo {author} {\bibfnamefont {F.}~\bibnamefont
  {Grill}}, \bibinfo {author} {\bibfnamefont {H.}~\bibnamefont {Pais}},
  \bibinfo {author} {\bibfnamefont {C.~m.~c.}\ \bibnamefont {Provid\^encia}},
  \bibinfo {author} {\bibfnamefont {I.}~\bibnamefont {Vida\~na}}, \ and\
  \bibinfo {author} {\bibfnamefont {S.~S.}\ \bibnamefont {Avancini}},\ }\href
  {\doibase 10.1103/PhysRevC.90.045803} {\bibfield  {journal} {\bibinfo
  {journal} {Phys. Rev. C}\ }\textbf {\bibinfo {volume} {90}},\ \bibinfo
  {pages} {045803} (\bibinfo {year} {2014})}\BibitemShut {NoStop}%
\bibitem [{\citenamefont {{Tolos}}\ \emph {et~al.}(2017)\citenamefont
  {{Tolos}}, \citenamefont {{Centelles}},\ and\ \citenamefont
  {{Ramos}}}]{2017PASA...34...65T}%
  \BibitemOpen
  \bibfield  {author} {\bibinfo {author} {\bibfnamefont {L.}~\bibnamefont
  {{Tolos}}}, \bibinfo {author} {\bibfnamefont {M.}~\bibnamefont
  {{Centelles}}}, \ and\ \bibinfo {author} {\bibfnamefont {A.}~\bibnamefont
  {{Ramos}}},\ }\href {\doibase 10.1017/pasa.2017.60} {\bibfield  {journal}
  {\bibinfo  {journal} {Publications of the Astronomical Society of Australia}\
  }\textbf {\bibinfo {volume} {34}},\ \bibinfo {eid} {e065} (\bibinfo {year}
  {2017})},\ \Eprint {http://arxiv.org/abs/1708.08681} {arXiv:1708.08681
  [astro-ph.HE]} \BibitemShut {NoStop}%
\bibitem [{\citenamefont {Providência}\ \emph {et~al.}(2019)\citenamefont
  {Providência}, \citenamefont {Fortin}, \citenamefont {Pais},\ and\
  \citenamefont {Rabhi}}]{10.3389/fspas.2019.00013}%
  \BibitemOpen
  \bibfield  {author} {\bibinfo {author} {\bibfnamefont {C.}~\bibnamefont
  {Providência}}, \bibinfo {author} {\bibfnamefont {M.}~\bibnamefont
  {Fortin}}, \bibinfo {author} {\bibfnamefont {H.}~\bibnamefont {Pais}}, \ and\
  \bibinfo {author} {\bibfnamefont {A.}~\bibnamefont {Rabhi}},\ }\href
  {\doibase 10.3389/fspas.2019.00013} {\bibfield  {journal} {\bibinfo
  {journal} {Frontiers in Astronomy and Space Sciences}\ }\textbf {\bibinfo
  {volume} {6}} (\bibinfo {year} {2019}),\
  10.3389/fspas.2019.00013}\BibitemShut {NoStop}%
\bibitem [{\citenamefont {Pearson}\ \emph {et~al.}(2019)\citenamefont
  {Pearson}, \citenamefont {Chamel}, \citenamefont {Potekhin}, \citenamefont
  {Fantina}, \citenamefont {Ducoin}, \citenamefont {Dutta},\ and\ \citenamefont
  {Goriely}}]{10.1093/mnras/stz800}%
  \BibitemOpen
  \bibfield  {author} {\bibinfo {author} {\bibfnamefont {J.~M.}\ \bibnamefont
  {Pearson}}, \bibinfo {author} {\bibfnamefont {N.}~\bibnamefont {Chamel}},
  \bibinfo {author} {\bibfnamefont {A.~Y.}\ \bibnamefont {Potekhin}}, \bibinfo
  {author} {\bibfnamefont {A.~F.}\ \bibnamefont {Fantina}}, \bibinfo {author}
  {\bibfnamefont {C.}~\bibnamefont {Ducoin}}, \bibinfo {author} {\bibfnamefont
  {A.~K.}\ \bibnamefont {Dutta}}, \ and\ \bibinfo {author} {\bibfnamefont
  {S.}~\bibnamefont {Goriely}},\ }\href {\doibase 10.1093/mnras/stz800}
  {\bibfield  {journal} {\bibinfo  {journal} {Monthly Notices of the Royal
  Astronomical Society}\ }\textbf {\bibinfo {volume} {486}},\ \bibinfo {pages}
  {768} (\bibinfo {year} {2019})},\ \Eprint
  {http://arxiv.org/abs/https://academic.oup.com/mnras/article-pdf/486/1/768/28345283/stz800.pdf}
  {https://academic.oup.com/mnras/article-pdf/486/1/768/28345283/stz800.pdf}
  \BibitemShut {NoStop}%
\bibitem [{\citenamefont {{Typel}}\ \emph {et~al.}(2010)\citenamefont
  {{Typel}}, \citenamefont {{R{\"o}pke}}, \citenamefont {{Kl{\"a}hn}},
  \citenamefont {{Blaschke}},\ and\ \citenamefont
  {{Wolter}}}]{2010PhRvC..81a5803T}%
  \BibitemOpen
  \bibfield  {author} {\bibinfo {author} {\bibfnamefont {S.}~\bibnamefont
  {{Typel}}}, \bibinfo {author} {\bibfnamefont {G.}~\bibnamefont
  {{R{\"o}pke}}}, \bibinfo {author} {\bibfnamefont {T.}~\bibnamefont
  {{Kl{\"a}hn}}}, \bibinfo {author} {\bibfnamefont {D.}~\bibnamefont
  {{Blaschke}}}, \ and\ \bibinfo {author} {\bibfnamefont {H.~H.}\ \bibnamefont
  {{Wolter}}},\ }\href {\doibase 10.1103/PhysRevC.81.015803} {\bibfield
  {journal} {\bibinfo  {journal} {\prc}\ }\textbf {\bibinfo {volume} {81}},\
  \bibinfo {eid} {015803} (\bibinfo {year} {2010})},\ \Eprint
  {http://arxiv.org/abs/0908.2344} {arXiv:0908.2344 [nucl-th]} \BibitemShut
  {NoStop}%
\bibitem [{\citenamefont {Carreau}\ \emph {et~al.}(2019)\citenamefont
  {Carreau}, \citenamefont {Gulminelli},\ and\ \citenamefont
  {Margueron}}]{Carreau2019}%
  \BibitemOpen
  \bibfield  {author} {\bibinfo {author} {\bibfnamefont {T.}~\bibnamefont
  {Carreau}}, \bibinfo {author} {\bibfnamefont {F.}~\bibnamefont {Gulminelli}},
  \ and\ \bibinfo {author} {\bibfnamefont {J.}~\bibnamefont {Margueron}},\
  }\href {\doibase 10.1140/epja/i2019-12884-1} {\bibfield  {journal} {\bibinfo
  {journal} {The European Physical Journal A}\ }\textbf {\bibinfo {volume}
  {55}},\ \bibinfo {pages} {188} (\bibinfo {year} {2019})}\BibitemShut
  {NoStop}%
\bibitem [{\citenamefont {Fortin}\ \emph {et~al.}(2016)\citenamefont {Fortin},
  \citenamefont {Provid\^encia}, \citenamefont {Raduta}, \citenamefont
  {Gulminelli}, \citenamefont {Zdunik}, \citenamefont {Haensel},\ and\
  \citenamefont {Bejger}}]{PhysRevC.94.035804}%
  \BibitemOpen
  \bibfield  {author} {\bibinfo {author} {\bibfnamefont {M.}~\bibnamefont
  {Fortin}}, \bibinfo {author} {\bibfnamefont {C.}~\bibnamefont
  {Provid\^encia}}, \bibinfo {author} {\bibfnamefont {A.~R.}\ \bibnamefont
  {Raduta}}, \bibinfo {author} {\bibfnamefont {F.}~\bibnamefont {Gulminelli}},
  \bibinfo {author} {\bibfnamefont {J.~L.}\ \bibnamefont {Zdunik}}, \bibinfo
  {author} {\bibfnamefont {P.}~\bibnamefont {Haensel}}, \ and\ \bibinfo
  {author} {\bibfnamefont {M.}~\bibnamefont {Bejger}},\ }\href {\doibase
  10.1103/PhysRevC.94.035804} {\bibfield  {journal} {\bibinfo  {journal} {Phys.
  Rev. C}\ }\textbf {\bibinfo {volume} {94}},\ \bibinfo {pages} {035804}
  (\bibinfo {year} {2016})}\BibitemShut {NoStop}%
\bibitem [{\citenamefont {Dinh~Thi}\ \emph {et~al.}(2021)\citenamefont
  {Dinh~Thi}, \citenamefont {Mondal},\ and\ \citenamefont
  {Gulminelli}}]{universe7100373}%
  \BibitemOpen
  \bibfield  {author} {\bibinfo {author} {\bibfnamefont {H.}~\bibnamefont
  {Dinh~Thi}}, \bibinfo {author} {\bibfnamefont {C.}~\bibnamefont {Mondal}}, \
  and\ \bibinfo {author} {\bibfnamefont {F.}~\bibnamefont {Gulminelli}},\
  }\href {\doibase 10.3390/universe7100373} {\bibfield  {journal} {\bibinfo
  {journal} {Universe}\ }\textbf {\bibinfo {volume} {7}} (\bibinfo {year}
  {2021}),\ 10.3390/universe7100373}\BibitemShut {NoStop}%
\bibitem [{\citenamefont {Read}\ \emph {et~al.}(2009)\citenamefont {Read},
  \citenamefont {Lackey}, \citenamefont {Owen},\ and\ \citenamefont
  {Friedman}}]{Read}%
  \BibitemOpen
  \bibfield  {author} {\bibinfo {author} {\bibfnamefont {J.~S.}\ \bibnamefont
  {Read}}, \bibinfo {author} {\bibfnamefont {B.~D.}\ \bibnamefont {Lackey}},
  \bibinfo {author} {\bibfnamefont {B.~J.}\ \bibnamefont {Owen}}, \ and\
  \bibinfo {author} {\bibfnamefont {J.~L.}\ \bibnamefont {Friedman}},\ }\href
  {\doibase 10.1103/PhysRevD.79.124032} {\bibfield  {journal} {\bibinfo
  {journal} {Phys. Rev. D}\ }\textbf {\bibinfo {volume} {79}},\ \bibinfo
  {pages} {124032} (\bibinfo {year} {2009})},\ \Eprint
  {http://arxiv.org/abs/0812.2163} {arXiv:0812.2163 [astro-ph]} \BibitemShut
  {NoStop}%
\bibitem [{\citenamefont {Raithel}\ \emph {et~al.}(2016)\citenamefont
  {Raithel}, \citenamefont {Ozel},\ and\ \citenamefont {Psaltis}}]{Raithel}%
  \BibitemOpen
  \bibfield  {author} {\bibinfo {author} {\bibfnamefont {C.~A.}\ \bibnamefont
  {Raithel}}, \bibinfo {author} {\bibfnamefont {F.}~\bibnamefont {Ozel}}, \
  and\ \bibinfo {author} {\bibfnamefont {D.}~\bibnamefont {Psaltis}},\ }\href
  {\doibase 10.3847/0004-637X/831/1/44} {\bibfield  {journal} {\bibinfo
  {journal} {Astrophys. J.}\ }\textbf {\bibinfo {volume} {831}},\ \bibinfo
  {pages} {44} (\bibinfo {year} {2016})},\ \Eprint
  {http://arxiv.org/abs/1605.03591} {arXiv:1605.03591 [astro-ph.HE]}
  \BibitemShut {NoStop}%
\bibitem [{\citenamefont {Suleiman}\ \emph {et~al.}(2022)\citenamefont
  {Suleiman}, \citenamefont {Fortin}, \citenamefont {Zdunik},\ and\
  \citenamefont {Providencia}}]{Suleiman}%
  \BibitemOpen
  \bibfield  {author} {\bibinfo {author} {\bibfnamefont {L.}~\bibnamefont
  {Suleiman}}, \bibinfo {author} {\bibfnamefont {M.}~\bibnamefont {Fortin}},
  \bibinfo {author} {\bibfnamefont {J.~L.}\ \bibnamefont {Zdunik}}, \ and\
  \bibinfo {author} {\bibfnamefont {C.}~\bibnamefont {Providencia}},\ }\href
  {\doibase 10.1103/PhysRevC.106.035805} {\bibfield  {journal} {\bibinfo
  {journal} {Phys. Rev. C}\ }\textbf {\bibinfo {volume} {106}},\ \bibinfo
  {pages} {035805} (\bibinfo {year} {2022})},\ \Eprint
  {http://arxiv.org/abs/2209.06052} {arXiv:2209.06052 [nucl-th]} \BibitemShut
  {NoStop}%
\bibitem [{\citenamefont {Lindblom}(2024)}]{Lindblom}%
  \BibitemOpen
  \bibfield  {author} {\bibinfo {author} {\bibfnamefont {L.}~\bibnamefont
  {Lindblom}},\ }\href {\doibase 10.1103/PhysRevD.110.043018} {\bibfield
  {journal} {\bibinfo  {journal} {Phys. Rev. D}\ }\textbf {\bibinfo {volume}
  {110}},\ \bibinfo {pages} {043018} (\bibinfo {year} {2024})},\ \Eprint
  {http://arxiv.org/abs/2407.16078} {arXiv:2407.16078 [astro-ph.HE]}
  \BibitemShut {NoStop}%
\bibitem [{\citenamefont {Reed}\ \emph {et~al.}(2024)\citenamefont {Reed},
  \citenamefont {Somasundaram}, \citenamefont {De}, \citenamefont {Armstrong},
  \citenamefont {Giuliani}, \citenamefont {Capano}, \citenamefont {Brown},\
  and\ \citenamefont {Tews}}]{Reed}%
  \BibitemOpen
  \bibfield  {author} {\bibinfo {author} {\bibfnamefont {B.~T.}\ \bibnamefont
  {Reed}}, \bibinfo {author} {\bibfnamefont {R.}~\bibnamefont {Somasundaram}},
  \bibinfo {author} {\bibfnamefont {S.}~\bibnamefont {De}}, \bibinfo {author}
  {\bibfnamefont {C.~L.}\ \bibnamefont {Armstrong}}, \bibinfo {author}
  {\bibfnamefont {P.}~\bibnamefont {Giuliani}}, \bibinfo {author}
  {\bibfnamefont {C.}~\bibnamefont {Capano}}, \bibinfo {author} {\bibfnamefont
  {D.~A.}\ \bibnamefont {Brown}}, \ and\ \bibinfo {author} {\bibfnamefont
  {I.}~\bibnamefont {Tews}},\ }\href@noop {} {\enquote {\bibinfo {title}
  {{Towards accelerated nuclear-physics parameter estimation from binary
  neutron star mergers: Emulators for the Tolman-Oppenheimer-Volkoff
  equations}},}\ } (\bibinfo {year} {2024}),\ \Eprint
  {http://arxiv.org/abs/2405.20558} {arXiv:2405.20558 [astro-ph.HE]}
  \BibitemShut {NoStop}%
\bibitem [{\citenamefont {{Fonseca}}\ \emph {et~al.}(2021)\citenamefont
  {{Fonseca}}, \citenamefont {{Cromartie}}, \citenamefont {{Pennucci}},
  \citenamefont {{Ray}}, \citenamefont {{Kirichenko}}, \citenamefont
  {{Ransom}}, \citenamefont {{Demorest}}, \citenamefont {{Stairs}},
  \citenamefont {{Arzoumanian}}, \citenamefont {{Guillemot}}, \citenamefont
  {{Parthasarathy}}, \citenamefont {{Kerr}}, \citenamefont {{Cognard}},
  \citenamefont {{Baker}}, \citenamefont {{Blumer}}, \citenamefont {{Brook}},
  \citenamefont {{DeCesar}}, \citenamefont {{Dolch}}, \citenamefont {{Dong}},
  \citenamefont {{Ferrara}}, \citenamefont {{Fiore}}, \citenamefont
  {{Garver-Daniels}}, \citenamefont {{Good}}, \citenamefont {{Jennings}},
  \citenamefont {{Jones}}, \citenamefont {{Kaspi}}, \citenamefont {{Lam}},
  \citenamefont {{Lorimer}}, \citenamefont {{Luo}}, \citenamefont {{McEwen}},
  \citenamefont {{McKee}}, \citenamefont {{McLaughlin}}, \citenamefont
  {{McMann}}, \citenamefont {{Meyers}}, \citenamefont {{Naidu}}, \citenamefont
  {{Ng}}, \citenamefont {{Nice}}, \citenamefont {{Pol}}, \citenamefont
  {{Radovan}}, \citenamefont {{Shapiro-Albert}}, \citenamefont {{Tan}},
  \citenamefont {{Tendulkar}}, \citenamefont {{Swiggum}}, \citenamefont
  {{Wahl}},\ and\ \citenamefont {{Zhu}}}]{2021ApJ...915L..12F}%
  \BibitemOpen
  \bibfield  {author} {\bibinfo {author} {\bibfnamefont {E.}~\bibnamefont
  {{Fonseca}}}, \bibinfo {author} {\bibfnamefont {H.~T.}\ \bibnamefont
  {{Cromartie}}}, \bibinfo {author} {\bibfnamefont {T.~T.}\ \bibnamefont
  {{Pennucci}}}, \bibinfo {author} {\bibfnamefont {P.~S.}\ \bibnamefont
  {{Ray}}}, \bibinfo {author} {\bibfnamefont {A.~Y.}\ \bibnamefont
  {{Kirichenko}}}, \bibinfo {author} {\bibfnamefont {S.~M.}\ \bibnamefont
  {{Ransom}}}, \bibinfo {author} {\bibfnamefont {P.~B.}\ \bibnamefont
  {{Demorest}}}, \bibinfo {author} {\bibfnamefont {I.~H.}\ \bibnamefont
  {{Stairs}}}, \bibinfo {author} {\bibfnamefont {Z.}~\bibnamefont
  {{Arzoumanian}}}, \bibinfo {author} {\bibfnamefont {L.}~\bibnamefont
  {{Guillemot}}}, \bibinfo {author} {\bibfnamefont {A.}~\bibnamefont
  {{Parthasarathy}}}, \bibinfo {author} {\bibfnamefont {M.}~\bibnamefont
  {{Kerr}}}, \bibinfo {author} {\bibfnamefont {I.}~\bibnamefont {{Cognard}}},
  \bibinfo {author} {\bibfnamefont {P.~T.}\ \bibnamefont {{Baker}}}, \bibinfo
  {author} {\bibfnamefont {H.}~\bibnamefont {{Blumer}}}, \bibinfo {author}
  {\bibfnamefont {P.~R.}\ \bibnamefont {{Brook}}}, \bibinfo {author}
  {\bibfnamefont {M.}~\bibnamefont {{DeCesar}}}, \bibinfo {author}
  {\bibfnamefont {T.}~\bibnamefont {{Dolch}}}, \bibinfo {author} {\bibfnamefont
  {F.~A.}\ \bibnamefont {{Dong}}}, \bibinfo {author} {\bibfnamefont {E.~C.}\
  \bibnamefont {{Ferrara}}}, \bibinfo {author} {\bibfnamefont {W.}~\bibnamefont
  {{Fiore}}}, \bibinfo {author} {\bibfnamefont {N.}~\bibnamefont
  {{Garver-Daniels}}}, \bibinfo {author} {\bibfnamefont {D.~C.}\ \bibnamefont
  {{Good}}}, \bibinfo {author} {\bibfnamefont {R.}~\bibnamefont {{Jennings}}},
  \bibinfo {author} {\bibfnamefont {M.~L.}\ \bibnamefont {{Jones}}}, \bibinfo
  {author} {\bibfnamefont {V.~M.}\ \bibnamefont {{Kaspi}}}, \bibinfo {author}
  {\bibfnamefont {M.~T.}\ \bibnamefont {{Lam}}}, \bibinfo {author}
  {\bibfnamefont {D.~R.}\ \bibnamefont {{Lorimer}}}, \bibinfo {author}
  {\bibfnamefont {J.}~\bibnamefont {{Luo}}}, \bibinfo {author} {\bibfnamefont
  {A.}~\bibnamefont {{McEwen}}}, \bibinfo {author} {\bibfnamefont {J.~W.}\
  \bibnamefont {{McKee}}}, \bibinfo {author} {\bibfnamefont {M.~A.}\
  \bibnamefont {{McLaughlin}}}, \bibinfo {author} {\bibfnamefont
  {N.}~\bibnamefont {{McMann}}}, \bibinfo {author} {\bibfnamefont {B.~W.}\
  \bibnamefont {{Meyers}}}, \bibinfo {author} {\bibfnamefont {A.}~\bibnamefont
  {{Naidu}}}, \bibinfo {author} {\bibfnamefont {C.}~\bibnamefont {{Ng}}},
  \bibinfo {author} {\bibfnamefont {D.~J.}\ \bibnamefont {{Nice}}}, \bibinfo
  {author} {\bibfnamefont {N.}~\bibnamefont {{Pol}}}, \bibinfo {author}
  {\bibfnamefont {H.~A.}\ \bibnamefont {{Radovan}}}, \bibinfo {author}
  {\bibfnamefont {B.}~\bibnamefont {{Shapiro-Albert}}}, \bibinfo {author}
  {\bibfnamefont {C.~M.}\ \bibnamefont {{Tan}}}, \bibinfo {author}
  {\bibfnamefont {S.~P.}\ \bibnamefont {{Tendulkar}}}, \bibinfo {author}
  {\bibfnamefont {J.~K.}\ \bibnamefont {{Swiggum}}}, \bibinfo {author}
  {\bibfnamefont {H.~M.}\ \bibnamefont {{Wahl}}}, \ and\ \bibinfo {author}
  {\bibfnamefont {W.~W.}\ \bibnamefont {{Zhu}}},\ }\href {\doibase
  10.3847/2041-8213/ac03b8} {\bibfield  {journal} {\bibinfo  {journal}
  {Astrophysical Journal, Letters}\ }\textbf {\bibinfo {volume} {915}},\
  \bibinfo {eid} {L12} (\bibinfo {year} {2021})},\ \Eprint
  {http://arxiv.org/abs/2104.00880} {arXiv:2104.00880 [astro-ph.HE]}
  \BibitemShut {NoStop}%
\bibitem [{\citenamefont {{Antoniadis}}\ \emph {et~al.}(2013)\citenamefont
  {{Antoniadis}}, \citenamefont {{Freire}}, \citenamefont {{Wex}},
  \citenamefont {{Tauris}}, \citenamefont {{Lynch}}, \citenamefont {{van
  Kerkwijk}}, \citenamefont {{Kramer}}, \citenamefont {{Bassa}}, \citenamefont
  {{Dhillon}}, \citenamefont {{Driebe}}, \citenamefont {{Hessels}},
  \citenamefont {{Kaspi}}, \citenamefont {{Kondratiev}}, \citenamefont
  {{Langer}}, \citenamefont {{Marsh}}, \citenamefont {{McLaughlin}},
  \citenamefont {{Pennucci}}, \citenamefont {{Ransom}}, \citenamefont
  {{Stairs}}, \citenamefont {{van Leeuwen}}, \citenamefont {{Verbiest}},\ and\
  \citenamefont {{Whelan}}}]{2013Sci...340..448A}%
  \BibitemOpen
  \bibfield  {author} {\bibinfo {author} {\bibfnamefont {J.}~\bibnamefont
  {{Antoniadis}}}, \bibinfo {author} {\bibfnamefont {P.~C.~C.}\ \bibnamefont
  {{Freire}}}, \bibinfo {author} {\bibfnamefont {N.}~\bibnamefont {{Wex}}},
  \bibinfo {author} {\bibfnamefont {T.~M.}\ \bibnamefont {{Tauris}}}, \bibinfo
  {author} {\bibfnamefont {R.~S.}\ \bibnamefont {{Lynch}}}, \bibinfo {author}
  {\bibfnamefont {M.~H.}\ \bibnamefont {{van Kerkwijk}}}, \bibinfo {author}
  {\bibfnamefont {M.}~\bibnamefont {{Kramer}}}, \bibinfo {author}
  {\bibfnamefont {C.}~\bibnamefont {{Bassa}}}, \bibinfo {author} {\bibfnamefont
  {V.~S.}\ \bibnamefont {{Dhillon}}}, \bibinfo {author} {\bibfnamefont
  {T.}~\bibnamefont {{Driebe}}}, \bibinfo {author} {\bibfnamefont {J.~W.~T.}\
  \bibnamefont {{Hessels}}}, \bibinfo {author} {\bibfnamefont {V.~M.}\
  \bibnamefont {{Kaspi}}}, \bibinfo {author} {\bibfnamefont {V.~I.}\
  \bibnamefont {{Kondratiev}}}, \bibinfo {author} {\bibfnamefont
  {N.}~\bibnamefont {{Langer}}}, \bibinfo {author} {\bibfnamefont {T.~R.}\
  \bibnamefont {{Marsh}}}, \bibinfo {author} {\bibfnamefont {M.~A.}\
  \bibnamefont {{McLaughlin}}}, \bibinfo {author} {\bibfnamefont {T.~T.}\
  \bibnamefont {{Pennucci}}}, \bibinfo {author} {\bibfnamefont {S.~M.}\
  \bibnamefont {{Ransom}}}, \bibinfo {author} {\bibfnamefont {I.~H.}\
  \bibnamefont {{Stairs}}}, \bibinfo {author} {\bibfnamefont {J.}~\bibnamefont
  {{van Leeuwen}}}, \bibinfo {author} {\bibfnamefont {J.~P.~W.}\ \bibnamefont
  {{Verbiest}}}, \ and\ \bibinfo {author} {\bibfnamefont {D.~G.}\ \bibnamefont
  {{Whelan}}},\ }\href {\doibase 10.1126/science.1233232} {\bibfield  {journal}
  {\bibinfo  {journal} {Science}\ }\textbf {\bibinfo {volume} {340}},\ \bibinfo
  {pages} {448} (\bibinfo {year} {2013})},\ \Eprint
  {http://arxiv.org/abs/1304.6875} {arXiv:1304.6875 [astro-ph.HE]} \BibitemShut
  {NoStop}%
\end{thebibliography}%
\end{document}